\documentclass[smallextended]{svjour3}       
\smartqed  
\usepackage{graphicx}
\usepackage{graphicx,times}
\usepackage{subfigure}
\usepackage{threeparttable}
\usepackage{booktabs}

\usepackage{enumerate}
\usepackage{amsmath}
\usepackage{cases}
\usepackage{longtable}
\usepackage{epstopdf}
\usepackage{amsmath,bm}
\usepackage{amssymb}
\usepackage{morefloats}
\usepackage{multirow}
\usepackage{array}
\usepackage{verbatim}
\begin{document}

\title{Small-scale turbulent dynamo in astrophysical environments: nonlinear dynamo and dynamo in a partially ionized plasma
\thanks{S.X. acknowledges the support for 
this work provided by NASA through the NASA Hubble Fellowship grant \# HST-HF2-51473.001-A awarded by the Space Telescope Science Institute, which is operated by the Association of Universities for Research in Astronomy, Incorporated, under NASA contract NAS5- 26555.
A.L. acknowledges the support of the NASA TCAN 144AAG1967.
The Flatiron Institute is supported by the Simons Foundation.
}
}


\author{Siyao Xu         \and
        Alex Lazarian 
}


\institute{Siyao Xu \at
              Institute for Advanced Study, 1 Einstein Drive, Princeton, NJ 08540, USA;
Hubble Fellow, \\
              \email{sxu@ias.edu}           
           \and
           Alex Lazarian \at
              Department of Astronomy, University of Wisconsin, 475 North Charter Street, Madison, WI 53706, USA; lazarian@astro.wisc.edu
              \at
              Center for Computation Astrophysics, Flatiron Institute, 162 5th Ave, New York, NY 10010, USA
}

\date{Received: date / Accepted: date}

\maketitle

\begin{abstract}
Small-scale turbulent dynamo is responsible for the amplification of magnetic fields on scales smaller than the driving scale of turbulence 
in diverse astrophysical media. 
Most earlier dynamo theories concern the kinematic regime and small-scale magnetic field amplification. 
Here we review our recent progress in developing the theories for the nonlinear dynamo and 
the dynamo regime in a partially ionized plasma.
The importance of reconnection diffusion of magnetic fields is identified for both the nonlinear dynamo 
and magnetic field amplification during gravitational contraction. 
For the dynamo in a partially ionized plasma, the coupling state between neutrals and ions and the ion-neutral collisional damping 
can significantly affect the dynamo behavior and the resulting magnetic field structure. 
We present both our analytical predictions and numerical tests with a two-fluid dynamo simulation 
on the dynamo features in this regime. 
In addition, to illustrate the astrophysical implications, we discuss several examples for the applications of the dynamo theory to studying 
magnetic field evolution in both preshock and postshock regions of supernova remnants, in weakly magnetized molecular clouds, 
during the (primordial) star formation, and during the first galaxy formation. 
\end{abstract}

\section{Introduction}
\label{intro}

Both turbulence and magnetic fields are ubiquitous in astrophysical plasmas, and 
they actively participate in diverse astrophysical processes, including the star formation, 
cosmic ray propagation, and magnetic reconnection and particle acceleration in high-energy astrophysical environments
\cite{Mckee_Ostriker2007,Brad13,Laz20}.
Turbulence is believed to be responsible for the amplification and maintenance of comic magnetic fields via the 
``small-scale" turbulent dynamo mechanism
\cite{Bran05}.
It refers to the amplification of magnetic fields due to the presence of turbulence 
on scales below the driving scale of turbulence. 
As the evidence for the magnetic field amplification via small-scale turbulent dynamo, 
the measured magnetic energy spectra in diverse astrophysical media over a vast range of length scales 
conform to the same spectral shape as the turbulent energy 
(e.g., \cite{Lea98,Chep98,Vog05}).
The dynamo amplification of magnetic fields and its implications have been studied 
in diverse contexts, including the large-scale structure formation 
\cite{Ryu08,Min15,Mar18,Domi19}, 
(primordial) star formation and galaxy formation 
\cite{Sur12,Schob13,Mc20,XuLgra20}.

In astrophysical reality, the evolution and properties of magnetic fields are determined by both  
the dynamo growth and diffusion of magnetic fields. 
Given drastically different conditions of astrophysical systems,
the turbulence parameters (e.g., driving scale, injected turbulent velocity) 
and plasma parameters (e.g., magnetic Prandtl number, {which is the ratio of viscosity to 
magnetic diffusivity (proportional to resistivity),}
ionization fraction) 
for describing the dynamo and diffusion processes 
can differ by many orders of magnitude between various systems 
\cite{Bran05}.
By contrast, numerical simulations of turbulent dynamo only cover a small range of turbulence and plasma parameters 
due to the limited numerical resolution 
\cite{Hau04}. 
Therefore, it is necessary to develop theories of turbulent dynamo applicable to 
different physical regimes and astrophysical conditions, 
which can also provide the recipe of the 
subgrid dynamo physics for large-scale simulations 
\cite{Mar18}.

Most earlier theoretical studies on turbulent dynamo focused on its kinematic regime with 
negligible magnetic back reaction
\cite{Batc50,Kaza68,KulA92,Subra98,Sch04}.
For the kinematic regime of dynamo, the dynamo growth of magnetic fields and the magnetic field structure are 
sensitive to the plasma parameters, 
including both the ionization fraction 
\cite{KulA92,Sub97}
and the magnetic Prandtl number
\cite{Bran19,Brand19}.
For astrophysical settings involving large length scales and long timescales, 
the kinematic dynamo alone is insufficient to explain the 
dynamo evolution of magnetic fields and their dynamical roles in the multi-scale astrophysical processes.
With the growth of magnetic energy, the magnetic back reaction on turbulence becomes significant 
and the dynamo enters the nonlinear regime.
Numerical simulations
reveal that the dynamo in the nonlinear regime has distinctive characteristics
\cite{Chod00,Hau03,Ryu08,CVB09,Bere11}.
Instead of an exponential growth of magnetic energy, 
the magnetic energy grows linearly with time with an inefficient dynamo growth rate. 
The resulting magnetic energy spectrum peaks at a large length scale instead of the resistive scale 
as for the kinematic dynamo 
\cite{Hau03,Bran05}.

Xu \& Lazarian (2016)
\cite{XL16}
(hereafter XL16)
developed an analytical theory of nonlinear turbulent dynamo and 
found the importance of reconnection diffusion (RD) of magnetic fields in a 
proper formulation of the nonlinear dynamo. 
RD of magnetic fields occurs for magnetic fields in the presence of turbulence. 
When magnetic fields are mixed by turbulence, 
the turbulent motion brings magnetic fields with different directions into contact and 
magnetic reconnection happens. 
This turbulence induced magnetic reconnection, i.e., turbulent reconnection, 
is an intrinsic part of magnetohydrodynamic (MHD) turbulence
\cite{LV99,Laz20}. 
It enables the RD of turbulent magnetic fields, which otherwise can only have wave-like oscillations
\cite{Laz05,Eyink2011,Ey15}.
{We note that RD is different from the turbulent diffusion that
is commonly used in mean-field dynamos. 
While the turbulent diffusion usually describes the diffusion of the weak magnetic field that can be treated as a passive vector, 
the RD describes the diffusion of the magnetic field that has a significant back reaction on turbulence.} 
In the nonlinear stage of dynamo, 
RD dominates over other microphysical diffusion effects arising from plasma processes, 
e.g., ambipolar diffusion, resistive diffusion, 
and it accounts for both the inefficient growth of magnetic energy 
and the large correlation scale of amplified magnetic fields
that cannot be explained by the kinematic dynamo.

Besides the nonlinear regime, XL16 also 
studied the kinematic dynamo at a large and small magnetic Prandtl number, 
and at different ionization fractions.
They found new regimes of kinematic dynamo, including 
the transitional stage in a high-Prandtl number medium with the shift of the correlation length of magnetic fields from 
the resistive scale to the viscous scale, 
and the damping stage of dynamo in a weakly ionized medium 
(see also \cite{Brand19}).
XL16 predictions on the dynamo behavior resulting from the 
weak coupling between neutrals and ions at a low ionization fraction 
have been numerical confirmed by the two-fluid dynamo simulation 
\cite{Xud19}.

{Turbulent dynamo naturally occurs when the turbulent energy exceeds the magnetic energy in astrophysical media, 
provided that there is a non-zero seed magnetic field. 
The seed field can be of primordial origin 
(\cite{Bierm50,Qua89,Laz92,Sigl97,Sch18}).} 
The XL16 theory has a broad range of astrophysical applications for studying the evolution and structure of magnetic fields, 
e.g., in supernova remnants and the implications on cosmic ray diffusion and acceleration
\cite{XuL17}, 
in gravitational collapse and primordial star formation 
\cite{Mc20,XuLgra20},
and in weakly ionized interstellar phases
\cite{XL17,Xud19}. 
In this review, we focus on the nonlinear turbulent dynamo (Section 2) and 
the dynamo in a partially ionized medium (Section 3) formulated in XL16. 
We also provide examples of their applications to studying magnetic field amplification in a variety of 
astrophysical processes and astrophysical environments
collected from our recent studies.

\section{Nonlinear turbulent dynamo}
\label{sec: noda}

Turbulent stretching of magnetic fields leads to the dynamo amplification of magnetic fields. 
We note that shock compression and gravitational compression in a contracting flow, as well as 
turbulent compression in compressible MHD turbulence, can also lead to amplification of magnetic fields. 
Here we distinguish the stretching of magnetic fields by turbulent shear from other mechanisms for magnetic field amplification, 
and only consider the former process as small-scale turbulent dynamo.

In the kinematic regime of turbulent dynamo with a negligible back reaction of magnetic fields, 
the dynamo theory was developed by 
\cite{Kaza68}
(see also \cite{Krai67,KulA92,Eyi10}).
When the magnetic back reaction becomes significant, we need to deal with the nonlinear turbulent dynamo. 
Nonlinear turbulent dynamo is characterized by the energy equipartition between turbulence and magnetic fields 
within the inertial range of turbulence.

\subsection{Theoretical formulation of nonlinear turbulent dynamo}

Magnetic reconnection in the presence of turbulence has the reconnection rate dependent on turbulence properties. 
This is known as the turbulent reconnection of magnetic fields
\cite{LV99}. 
Turbulent reconnection is an intrinsic part of MHD turbulent cascade and enables the turbulent motions of magnetic fields, 
which otherwise would only have wave-like oscillations. 
The turbulent reconnection theory has been tested both numerically and observationally, with a broad range of applications 
in a variety of astrophysical environments 
(see the recent review by \cite{Laz20}).

Diffusion of magnetic fields relative to plasma as a result of turbulent reconnection is termed as ``reconnection diffusion (RD)"
\cite{Laz05,Eyink2011,Ey15}.
It has been applied in the context of star formation and well explains observations 
(e.g.,\cite{Sant10,LEC12,Laz14r}). 
Despite its importance, RD was disregarded in earlier studies of nonlinear turbulent dynamo. 
We stress that although some microphysical effects, e.g., ambipolar diffusion, 
were discussed in the literature as nonlinear effects 
\cite{KulA92,Subra98}, 
the ``nonlinear" dynamo considered here is characterized by the energy equipartition between turbulence and magnetic fields 
and by the development of MHD turbulence because of the dynamical effect of magnetic fields on turbulence.

The dynamo growth rate is determined by both the turbulent stretching of magnetic fields and diffusion of magnetic fields, 
as the former amplifies magnetic fields, and the latter causes loss of magnetic fields. 
As another important characteristic of nonlinear dynamo,
RD of magnetic fields dominates over other diffusion effects 
arising from plasma microphysics, e.g., ambipolar diffusion (related to ionization fraction), resistive diffusion 
(related to magnetic Prandtl number).
Otherwise the MHD turbulence would be suppressed by microscopic diffusion effects on all length scales, 
and the dynamo would not be in the nonlinear regime. 
Therefore, the dynamo growth rate of nonlinear dynamo is only dependent on the parameters of turbulence.

XL16 derived the nonlinear dynamo theory and found that the magnetic energy $ \mathcal{E}$ grows linearly with time, 
\begin{equation}\label{eq: ennoncr}
    \mathcal{E} = \mathcal{E}_\text{cr} + \frac{3}{38} \epsilon (t-t_\text{cr}),
\end{equation}
where $\mathcal{E}_\text{cr}$ is the magnetic energy at the beginning of the nonlinear stage at the time $t = t_\text{cr}$. 
In a system with the initial magnetic energy $\mathcal{E}_\text{cr}$ in equipartition with the turbulent energy at some length 
scale within the inertial range of turbulence, that is, the turbulence is super-Alfv\'{e}nic (see Section \ref{ssec: mwmmc}), 
we have $t_\text{cr} = 0$. 
In a system with an initially very weak magnetic field, the turbulent dynamo is first in the kinematic regime until the energy 
equipartition between turbulence and magnetic field can be reached at some length scale within the inertial range. 
In this case, $t_\text{cr}$ is given by the duration of the kinematic stage and depends on the plasma conditions
(XL16).
The linear dependence of $\mathcal{E}$ on time comes from the constant energy transfer rate $\epsilon= v_k^3 k$ of turbulence. 
Here $v_k$ is the turbulent velocity corresponding to the wavenumber $k$.
The turbulent energy cascade with the turbulent kinetic energy 
transferred from 
one eddy to the next smaller eddy has a constant energy transfer rate.
Meanwhile, a small fraction $3/38$ of the turbulent energy is converted to magnetic energy within the eddy turnover time, 
resulting in a constant magnetic energy growth rate.  
The low efficiency of nonlinear dynamo growth is the consequence of reconnection diffusion. 
The growing magnetic energy is passed down toward smaller scales along the turbulent energy cascade. 
The above small growth rate of nonlinear dynamo is quantitatively consistent with that measured in nonlinear dynamo simulations by, e.g., 
\cite{Ryu08,CVB09}.

\begin{figure*}[ht]
\centering
\subfigure[$P_m=1$]{
   \includegraphics[width=8cm]{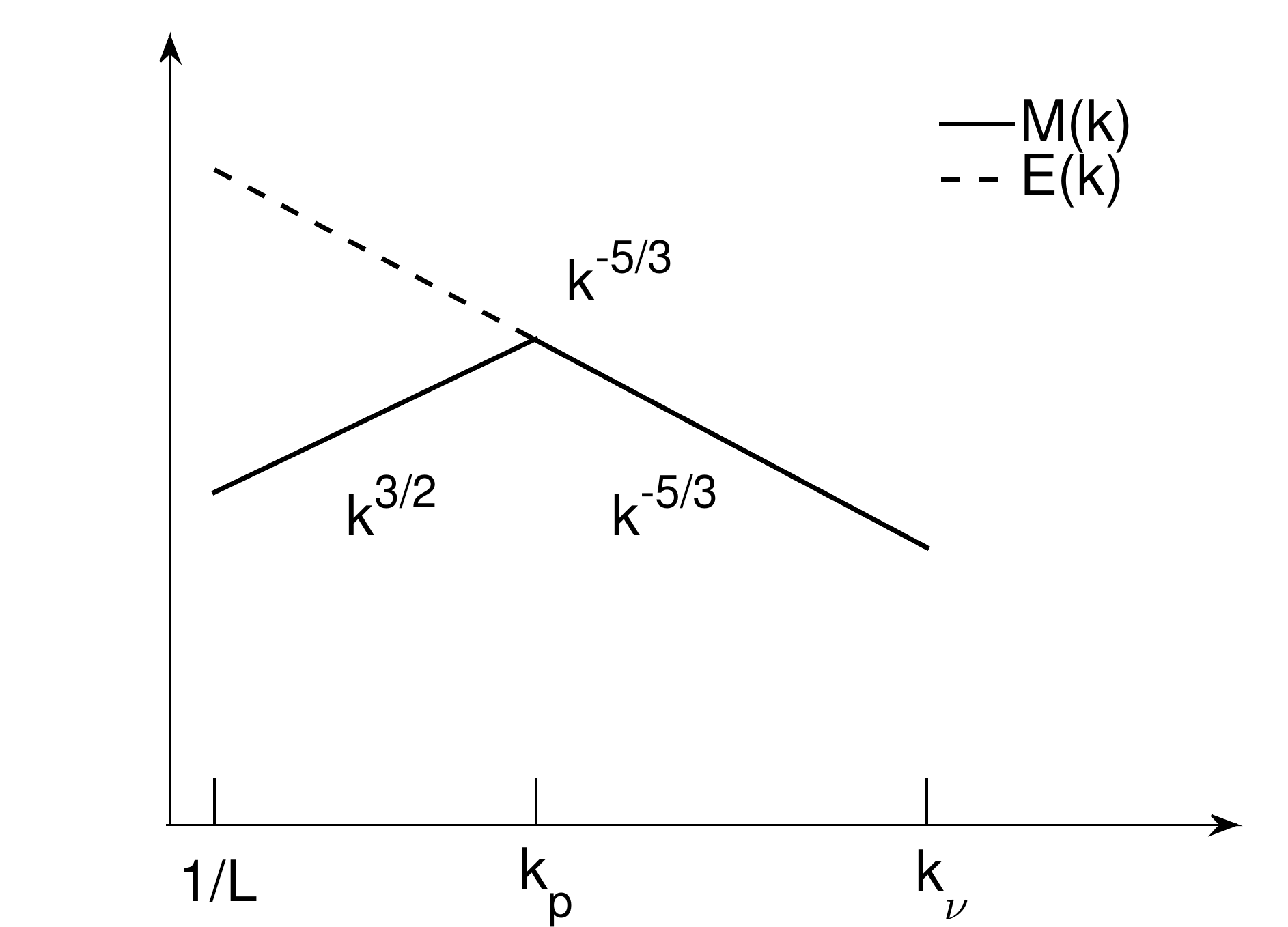}\label{fig: pmlt}}
\subfigure[$P_m\gg1$]{
   \includegraphics[width=8cm]{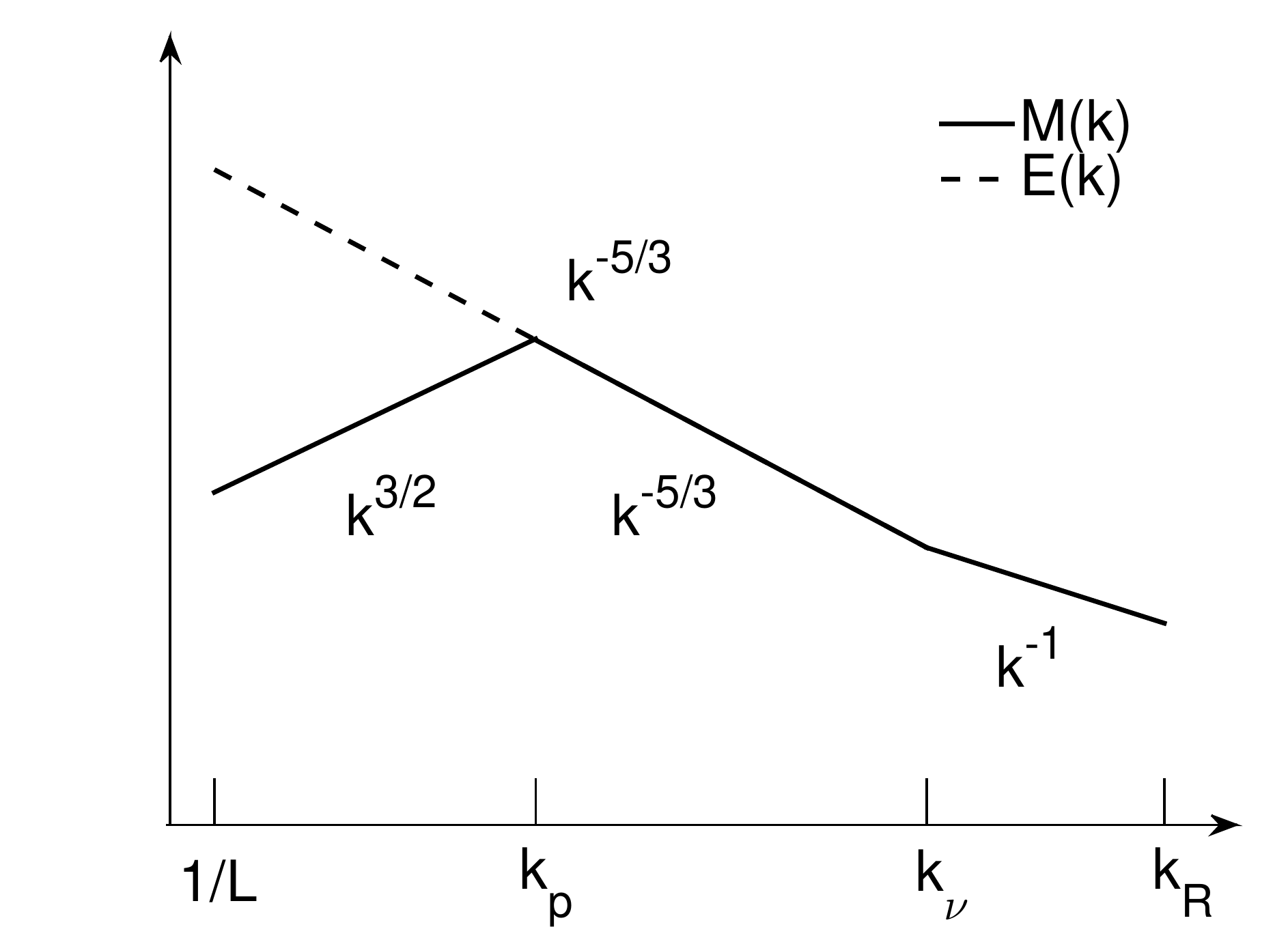}\label{fig: pmht}}
\subfigure[$P_m=1$, figure. 5.1 in \cite{Bran05}]{
   \includegraphics[width=8cm]{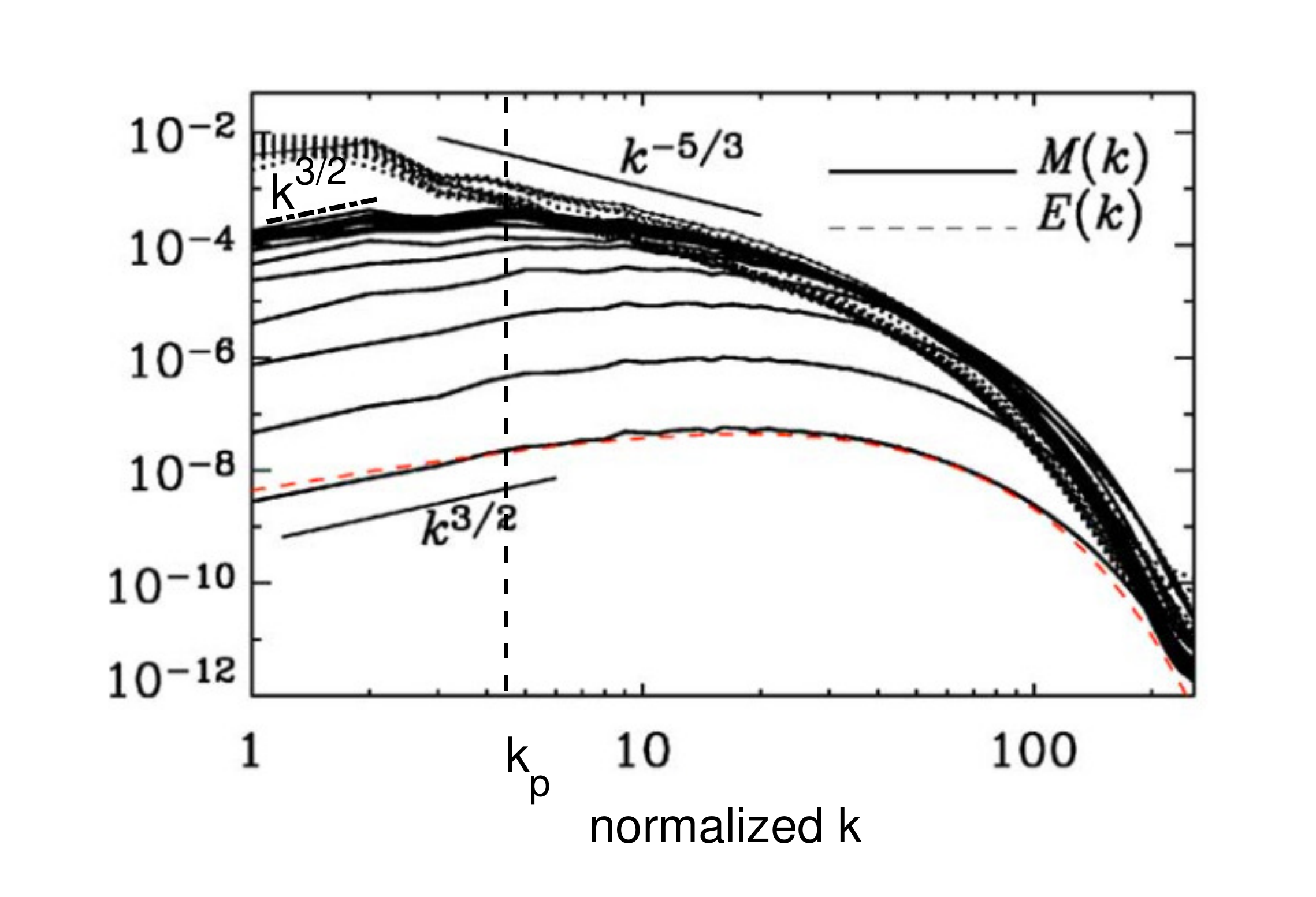}\label{fig: pmls}}
\subfigure[$P_m=50$, figure. 5.2 in \cite{Bran05}]{
   \includegraphics[width=8cm]{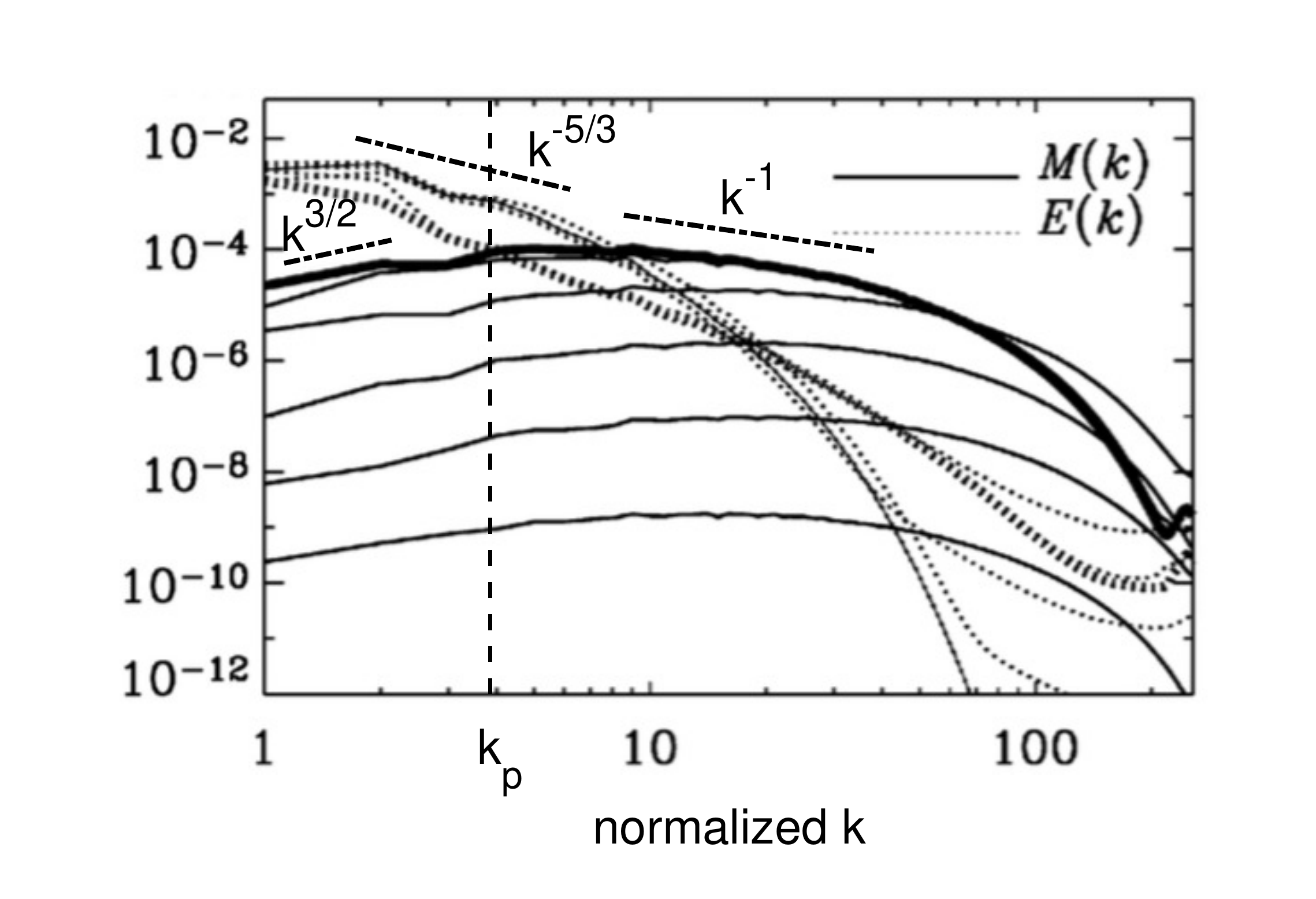}\label{fig: pmhs}}   
\caption{ Comparison between the analytical expectation (a,b) and numerical simulation (c,d) on the magnetic energy spectrum 
$M(k)$ during the nonlinear stage of dynamo at $P_m =1$ and $P_m\gg1$.
$E(k)$ is turbulent energy spectrum. 
From XL16.}
\label{fig: sket}
\end{figure*}

The correlation length $1/k_p$ of the stretched magnetic fields is given by the size of turbulent eddies that dominate the stretching. 
As the energy equipartition between turbulence and magnetic fields is reached scale by scale, the correlation length of magnetic fields 
increases with time (XL16), 
\begin{equation}\label{eq: evokpcri}
    k_p = \Big[k_\text{cr}^{-\frac{2}{3}} + \frac{3}{19} \epsilon^\frac{1}{3}(t-t_\text{cr})\Big]^{-\frac{3}{2}},
\end{equation}
where $k_\text{cr}$ is the correlation wavenumber of magnetic fields at $t = t_\text{cr}$. 
As an illustration, Fig. \ref{fig: sket} shows the magnetic energy spectrum $M(k)$ in the nonlinear stage of dynamo with magnetic Prandtl number 
$P_m = 1$ and $P_m \gg 1$. 
$1/k_p$ is the peak scale of $M(k)$. 
At $k>k_p$, magnetic energy is in equipartition with turbulent energy, and 
$M(k)$ follows the same Kolmogorov energy spectrum as that of hydrodynamic turbulence due to reconnection diffusion. 
At $k<k_p$, magnetic fields are relatively weak, and 
$M(k)$ has the form of Kazantsev spectrum as in the kinematic regime of dynamo. 
The analytical expectation in XL16 agrees well with the simulations by 
\cite{Bran05}. 
As we discussed, the nonlinear dynamo is independent of plasma effects and thus independent of $P_m$. 
The short inertial range of turbulence seen in the dynamo simulation at $P_m = 50$ is due to the limited numerical resolution. 

At the final saturation of nonlinear dynamo, magnetic energy is in approximate equipartition with the turbulent energy at the injection scale $L$, 
\begin{equation}\label{eq: satnlf}
    \mathcal{E}_\text{sat,nl} = \frac{1}{2} V_L^2,
\end{equation}
where $V_L$ is the turbulent velocity at $L$. 
The time required for the nonlinear dynamo to reach final saturation is approximately 
\begin{equation}\label{eq: nontdrl}
    \tau_\text{nl} \approx \frac{19}{3} \frac{L}{V_L}. 
\end{equation}
In the above expression, we assume that the turnover time of the smallest eddies are negligible compared with 
that of the largest eddy at $L$. 
The timescale of nonlinear dynamo is much longer than the turbulent eddy turnover time at $L$ 
because of its inefficient dynamo growth. 
The relatively long timescale of the nonlinear dynamo and its low efficiency 
indicate that only a small fraction of turbulent energy contributes to the growth of magnetic energy. 
This result is also consistent with that in dynamo simulations 
in a cosmological context
\cite{Ryu08,Min15,Domi19}
and in supernova remnants (see Section \ref{ssec: shpost}).

As the observational evidence for nonlinear dynamo in astrophysical media, 
the large correlation length of magnetic fields and Kolmogorov spectrum of magnetic energy are indicated by magnetic field 
measurements in the interstellar medium (ISM)
(e.g., \cite{Chep98})
and the intracluster medium (ICM)
(e.g., \cite{Vog05}).

\subsection{Magnetic field amplification in weakly magnetized molecular clouds}
\label{ssec: mwmmc}

Turbulent dynamo takes place when the turbulent energy is larger than the magnetic energy. 
In a super-Alfv\'{e}nic turbulent molecular cloud (MC) with the initial Alfv\'{e}nic Mach number $M_A = V_L / V_A$ larger than unity, 
where $V_A = B/ \sqrt{4 \pi \rho}$ is the Alfv\'{e}n speed, $B$ is the magnetic field strength, and $\rho$ is density,
nonlinear dynamo occurs within the inertial range of turbulence. 
Turbulence in MCs can originate from the global cascade of turbulence driven by supernova explosions 
\cite{Pad16}.

\begin{figure*}[htbp]
\centering
\subfigure[Initial state ]{
   \includegraphics[width=8cm]{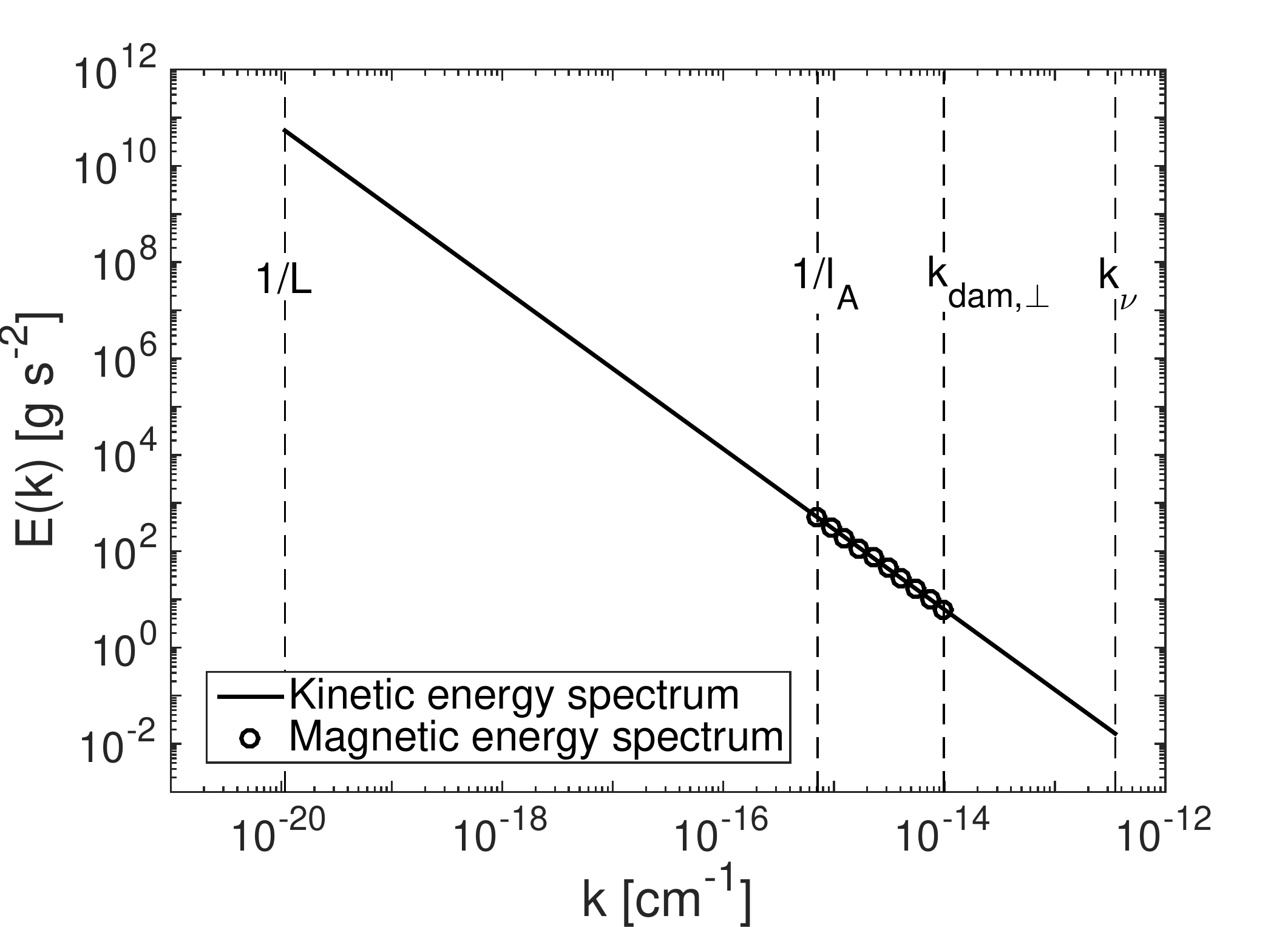}\label{fig: din}}
\subfigure[Final saturated state ]{
   \includegraphics[width=8cm]{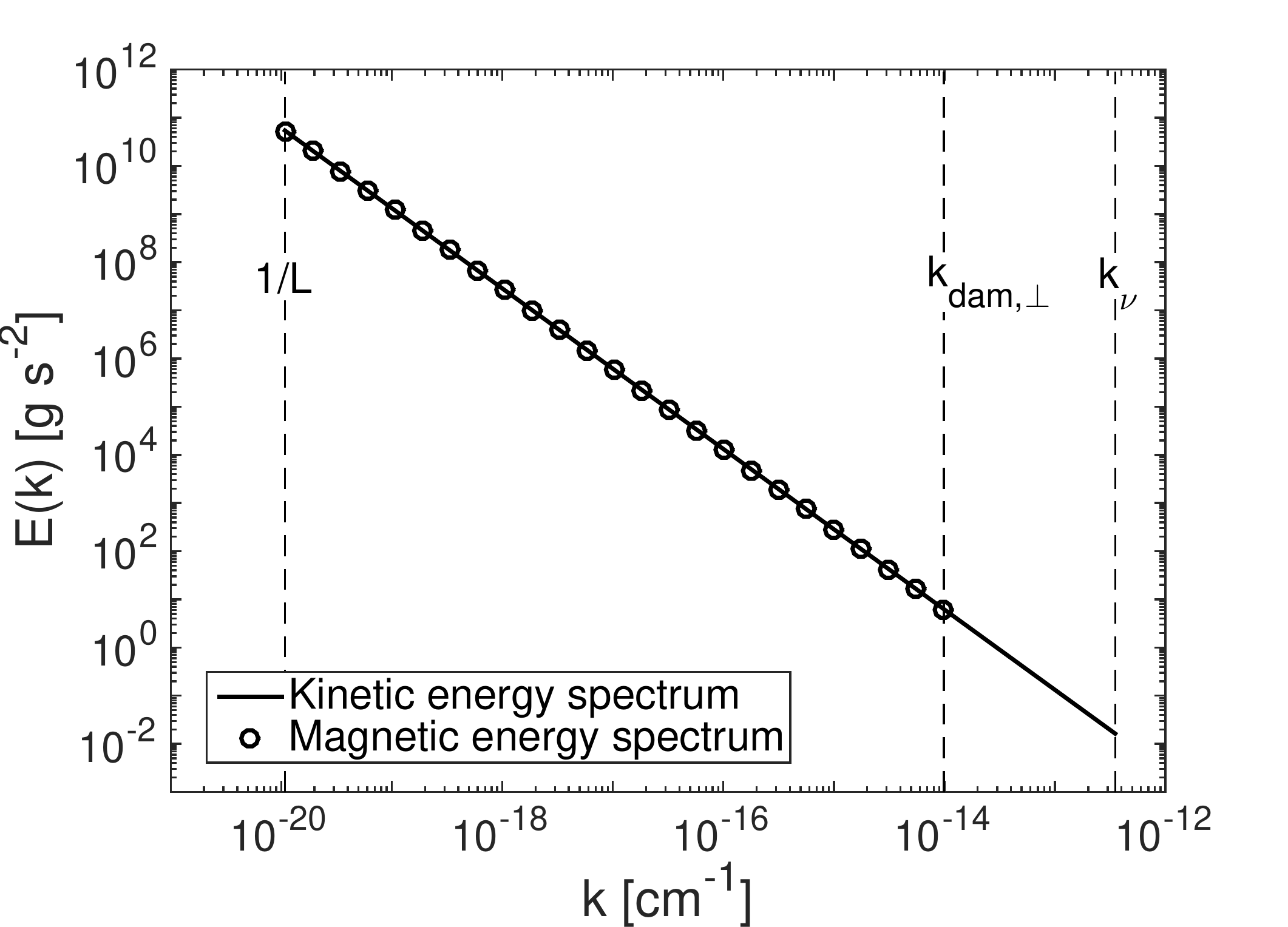}\label{fig: dfa}}  
\caption{ Magnetic energy spectrum at the initial state (a) and finial state (b) of the nonlinear turbulent dynamo in 
an initially super-Alfv\'{e}nic MC. 
From \cite{XL17}.}
\label{fig: dynmc}
\end{figure*}

As an illustration, Fig. \ref{fig: dynmc} shows the evolution of magnetic energy spectrum due to nonlinear turbulent dynamo in an 
initially weakly magnetized MC. It follows the same Kolmogorov form as the turbulent spectrum. 
$l_A$ corresponds to the energy equipartition scale and the peak scale of the evolving magnetic energy spectrum. 
It increases with time (Eq. \eqref{eq: evokpcri}) until reaching $l_A = L$ at the final saturation, and the turbulence becomes 
trans-Alfv\'{e}nic with $M_A=1$. 
$k_{\text{dam},\perp}$ is the perpendicular ion-neutral collisional damping wavenumber in a weakly ionized medium (see Section \ref{sec: part}), 
and $k_\nu$ is the viscous wavenumber due to neutral viscosity. 
The ion-neutral collisional damping usually dominates over the neutral viscous damping in a weakly ionized medium 
\cite{XLY14,Xuc16,XL17}.

We note that 
if the cloud lifetime is comparable to the turbulent crossing time
\cite{Elm00}, 
then as the inefficient nonlinear dynamo has the timescale much longer than the largest eddy turnover time (Eq. \eqref{eq: nontdrl}), 
final energy equipartition between turbulence and magnetic fields at $L$
may not be reached in an initially highly super-Alfv\'{e}nic MC. 
It means that the turbulence in the MC would stay super-Alfv\'{e}nic during the star formation process. 
This finding can have important implications for studying the role of magnetic fields in star formation 
\cite{Cru12}.

\subsection{Magnetic field amplification in supernova remnants: postshock region}
\label{ssec: shpost}

Strong magnetic fields in supernova remnants (SNRs) are inferred from observations 
(e.g., \cite{Lon94,Bamb05,Vin12})
and are also required for confining cosmic ray particles for acceleration. 
The above nonlinear dynamo theory has been applied to studying magnetic field amplification in the SNR postshock region in 
\cite{XuL17}. 
Different from the interstellar turbulence, the turbulence in the SNR postshock region is driven by the 
interaction between supernova shocks and interstellar density inhomogeneities. 
The driving scale is comparable to the size of density clumps in the upstream ISM.  

\begin{figure*}[ht]
\centering
\subfigure[Temporal evolution of $B$]{
   \includegraphics[width=8cm]{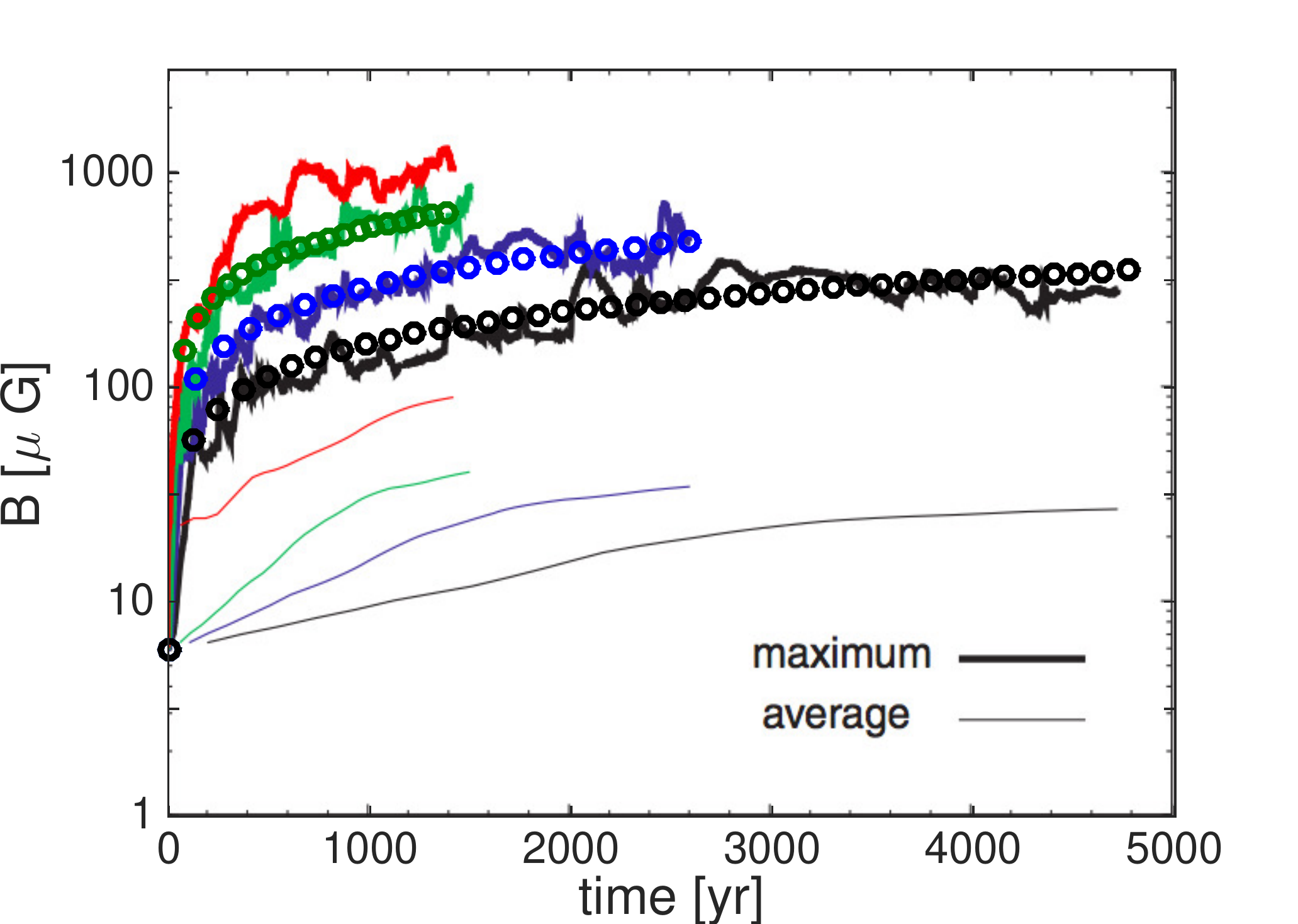}\label{fig: sh1}}
\subfigure[Spatial profiles of $B$]{
   \includegraphics[width=8cm]{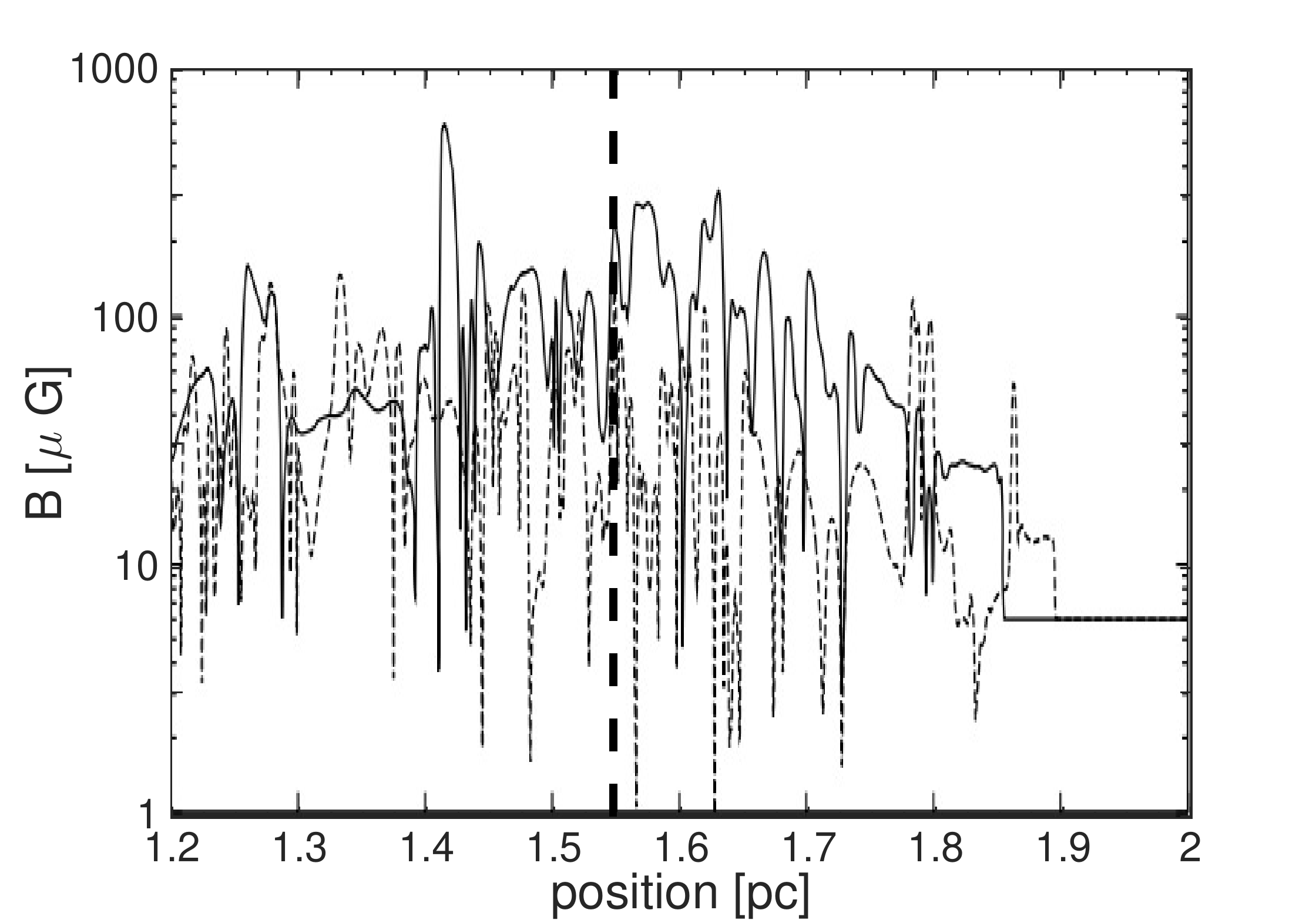}\label{fig: sh2}}  
\caption{ Comparisons between the nonlinear dynamo theory (XL16) and the numerical results in \cite{Ino09} for 
magnetic field amplification in the SNR postshock region. 
The circles in (a) and the 
vertical dashed line in (b) are analytical calculations. 
From \cite{XuL17}. 
}
\label{fig: dynsh}
\end{figure*}

In \cite{XuL17}, the magnetic field amplification based on the 
nonlinear dynamo theory is compared with that measured in shock simulations 
\cite{Ino09} (see Fig. \ref{fig: dynsh}). 
The linear-in-time growth of magnetic energy with a small dynamo growth rate as expected for nonlinear dynamo theory 
is seen from the simulated temporal evolution of magnetic fields (Fig. \ref{fig: sh1}). 
The maximum field strength is determined by the injected turbulent velocity, 
which depends on the shock velocity in different numerical models. 
The position where the nonlinear dynamo saturates and the magnetic field reaches its highest strength (indicated by the vertical dashed line
in Fig. \ref{fig: sh2})
is determined by 
the shock velocity and the nonlinear dynamo timescale. 
The long timescale of the inefficient nonlinear dynamo well explains the position of the maximum magnetic field, which is far 
away from the shock front.

Both results naturally explain the X-ray hot spots detected at more than $0.1$ pc in the SNR postshock region, which 
suggest the presence of magnetic field of $\sim1$ mG far behind the shock front 
\cite{Uch07,Uch08}.

\subsection{Magnetic field amplification during gravitational collapse}

In a weakly magnetized and gravitationally collapsing system, 
magnetic fields are amplified 
not only by turbulent dynamo, but also by gravitational compression. 
The growth of magnetic energy in this scenario can be described as 
\cite{XuLgra20}
\begin{equation}\label{eq: gelemen}
       \mathcal{E} =  \frac{3}{38}  \epsilon_0 C^\frac{4\alpha}{19} \int_{t_\text{cr}}^t C^{-\frac{3}{2} - \frac{4\alpha}{19}} dt  + \mathcal{E}_\text{cr} C^\frac{4\alpha}{19},
\end{equation}
where $C = r/ r_0$ is defined as the compression factor, $r_0$ is the initial radius of the collapsing sphere, and $r$ is the radius at time $t$. 
$\epsilon_0$ is the initial energy transfer rate of turbulence, 
and $\alpha =-1$ for isotropic compression. 
The above formula can recover the expression in Eq. \eqref{eq: ennoncr} at $C=1$ for nonlinear dynamo without gravitational compression. 
The correlation wavenumber of the amplified magnetic fields evolves as 
\cite{XuLgra20}
\begin{equation}
     k_p = \Bigg[\frac{3}{19}\epsilon_0^\frac{1}{3} C^{1+\frac{4 \alpha}{19}} \int_{t_\text{cr}}^t C^{-\frac{3}{2} - \frac{4\alpha}{19}} dt  + 2 \epsilon_0^{-\frac{2}{3}} \mathcal{E}_\text{cr} C^{1 + \frac{4\alpha}{19}}  \Bigg]^{-\frac{3}{2}}.
\end{equation}
The correlation length $1/k_p$
becomes comparable to the driving scale of turbulence at the full saturation of nonlinear dynamo.

\begin{figure*}[ht]
\centering   
\subfigure[]{
   \includegraphics[width=8.5cm]{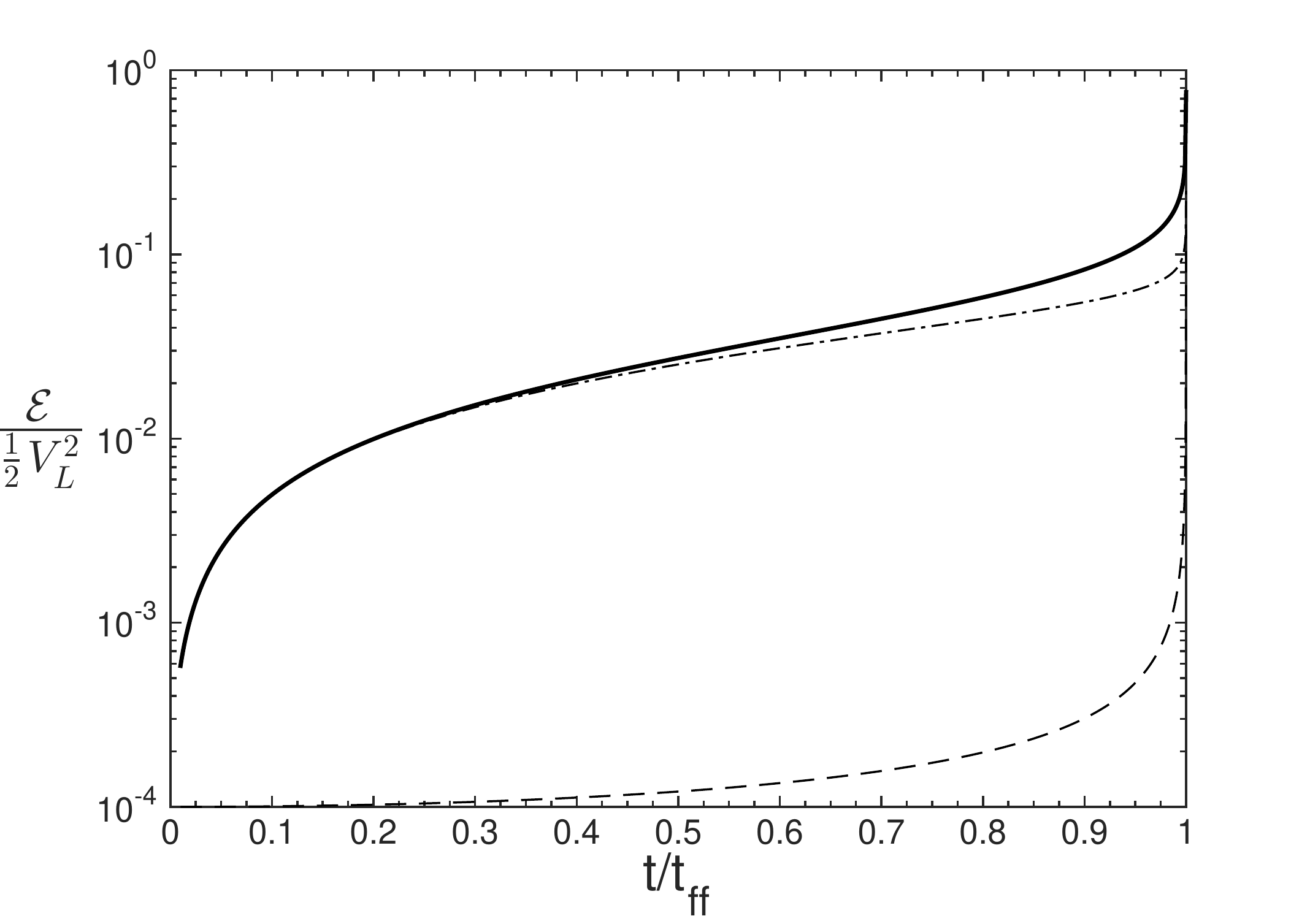}\label{fig: enetim}}
\subfigure[]{
   \includegraphics[width=8.5cm]{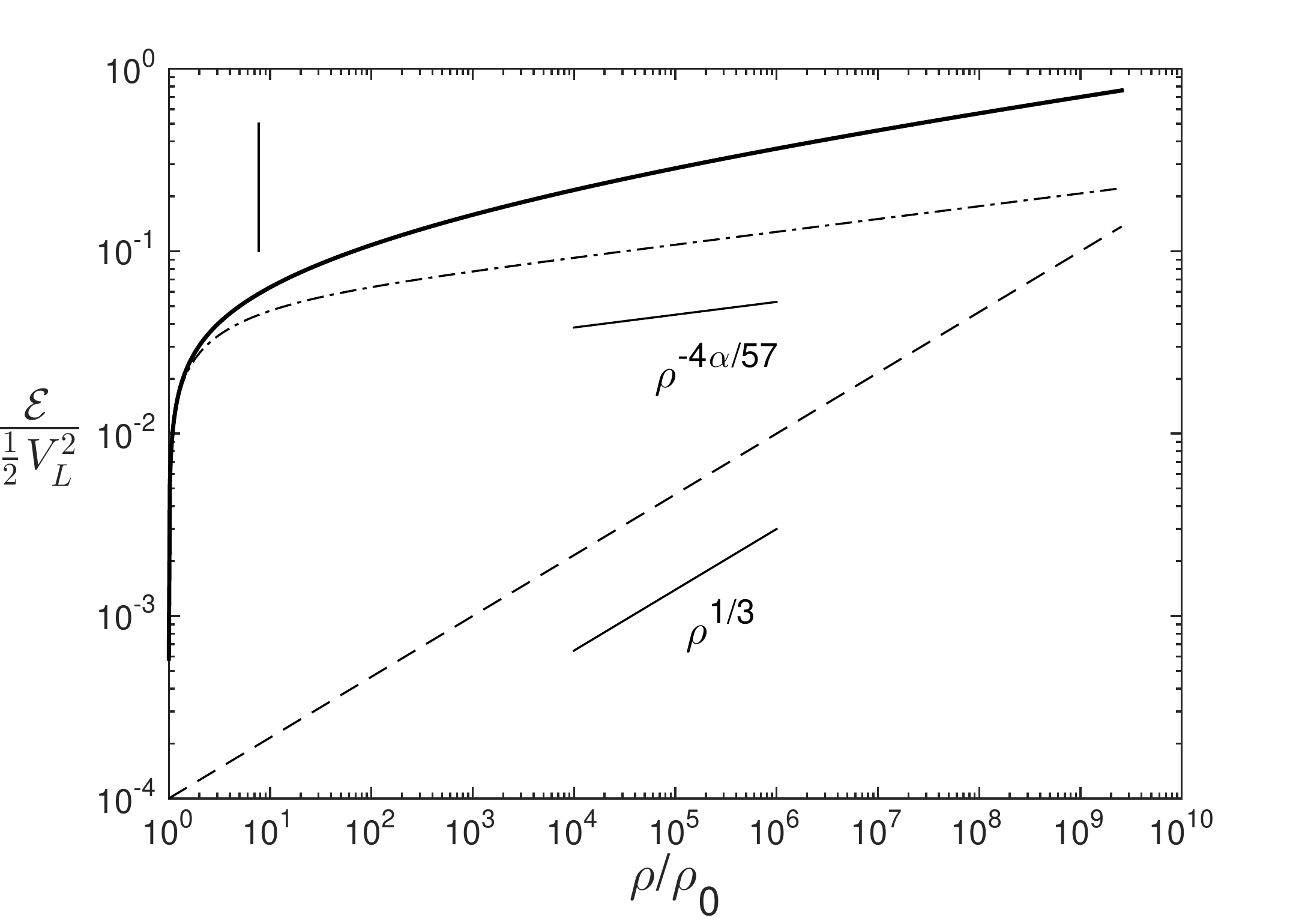}\label{fig: enerat}}
\caption{ Normalized $\mathcal{E}$ as a function of normalized $t$ in (a) and as a function of 
normalized $\rho$ in (b). 
The solid line is given by 
Eq. \eqref{eq: genumrat} when the reconnection diffusion of magnetic fields is considered. 
The dash-dotted line is its approximation given by 
Eq. \eqref{eq: ratsht}. 
The dashed line shows the flux-freezing scaling as a comparison 
(Eq. \eqref{eq: refffs}).
The short vertical line in (b) corresponds to $t/t_{ff} = 0.8$. 
From \cite{XuLgra20}.}
\label{fig: ene}
\end{figure*}

For the 
free-fall collapse of a uniform sphere 
\cite{Spit68},
we consider the compression factor as  
\cite{Giri14,Mc20},
\begin{equation}\label{eq: spitzc}
     C  \approx \Bigg(1- \Big(\frac{t}{t_{ff}}\Big)^2\Bigg)^\frac{2}{3} ,
\end{equation}
where $t_{ff}$ is the initial free-fall time of the sphere. 
In this case,
$\mathcal{E}$, normalized by the turbulent energy $V_L^2/2$ at $L$, evolves as 
\cite{XuLgra20}
\begin{align}
     \frac{\mathcal{E}}{ \frac{1}{2} V_L^2} &\approx \frac{3}{19} \sqrt{\frac{3}{32}} C^\frac{4\alpha}{19}  \int_{\frac{t_\text{cr}}{t_{ff}}}^{\frac{t}{t_{ff}}} \Bigg(1- \Big(\frac{t}{t_{ff}}\Big)^2\Bigg)^\beta d \Big(\frac{t}{t_{ff}}\Big)    \nonumber\\
     & ~~~~ + \frac{\mathcal{E}_\text{cr}}{\frac{1}{2} V_L^2} C^\frac{4\alpha}{19}   \label{eq: genumrat}\\
     & \approx \frac{3}{19} \sqrt{\frac{3}{32}} C^\frac{4\alpha}{19} \sqrt{1-C^\frac{3}{2}} + \frac{\mathcal{E}_\text{cr}}{\frac{1}{2} V_L^2} C^\frac{4\alpha}{19}  \label{eq: ratsht} ,
\end{align}
where Eq. \eqref{eq: ratsht} is its approximate form at a short time. 
Fig. \ref{fig: ene} shows the evolution of the normzlied $\mathcal{E}$ as a function of $t$ and density $\rho$,
where $\rho_0$ is the initial density. 
The growth of $\mathcal{E}$ is initially dominated by the nonlinear turbulent dynamo when the change of $\rho$ is insignificant. 
With the rapid growth of $\rho$ at a later time, 
the gravitational compression becomes the dominant effect on amplifying magnetic fields. 
The scaling becomes slightly steeper than 
\begin{equation}\label{eq: deerdw}
     \mathcal{E} \propto C^\frac{4\alpha}{19} \propto \rho^{-\frac{4\alpha}{57}}.
\end{equation}
The relation of $C$ to $\rho$ is given by 
\begin{equation}\label{eq: rrhcsp}
     \frac{\rho}{\rho_0}  = \frac{r_0^3}{r^3} = C^{-3}.
\end{equation}
We note that the slope $4\alpha/19$ is derived based on the 
Kolmogorov spectrum of turbulent energy 
and the Kazantsev spectrum of magnetic energy
(see also e.g., \cite{Krai67,Eyi10}
for a different form of Kazantsev spectrum), 
and is independent of the detailed model for collapse.

As a comparison with the flux-freezing scaling,
we also present the result in Fig. \ref{fig: ene} for compression alone under the freezing-in condition, 
\begin{equation}\label{eq: refffs}
    \frac{\mathcal{E}}{ \frac{1}{2} V_L^2}  = \frac{\mathcal{E}_\text{cr}}{ \frac{1}{2} V_L^2} C^\alpha.
\end{equation}
It grows steeply with $\rho$.
By contrast, when the reconnection diffusion of magnetic field is taken into account, 
$\mathcal{E}$ has a much weaker dependence on $\rho$.
The removal of magnetic flux during compression has been numerically demonstrated by, e.g, 
\cite{Sant10}.
Despite the additional magnetic field amplification via gravitational compression, 
the balance between the growth of magnetic fields and the reconnection diffusion of magnetic fields still stands, 
and the growth ceases when $\mathcal{E}$ reaches equipartition with the turbulent energy of the largest eddy at $L$.

\begin{figure}[htbp]
\centering   
   \includegraphics[width=8.5cm]{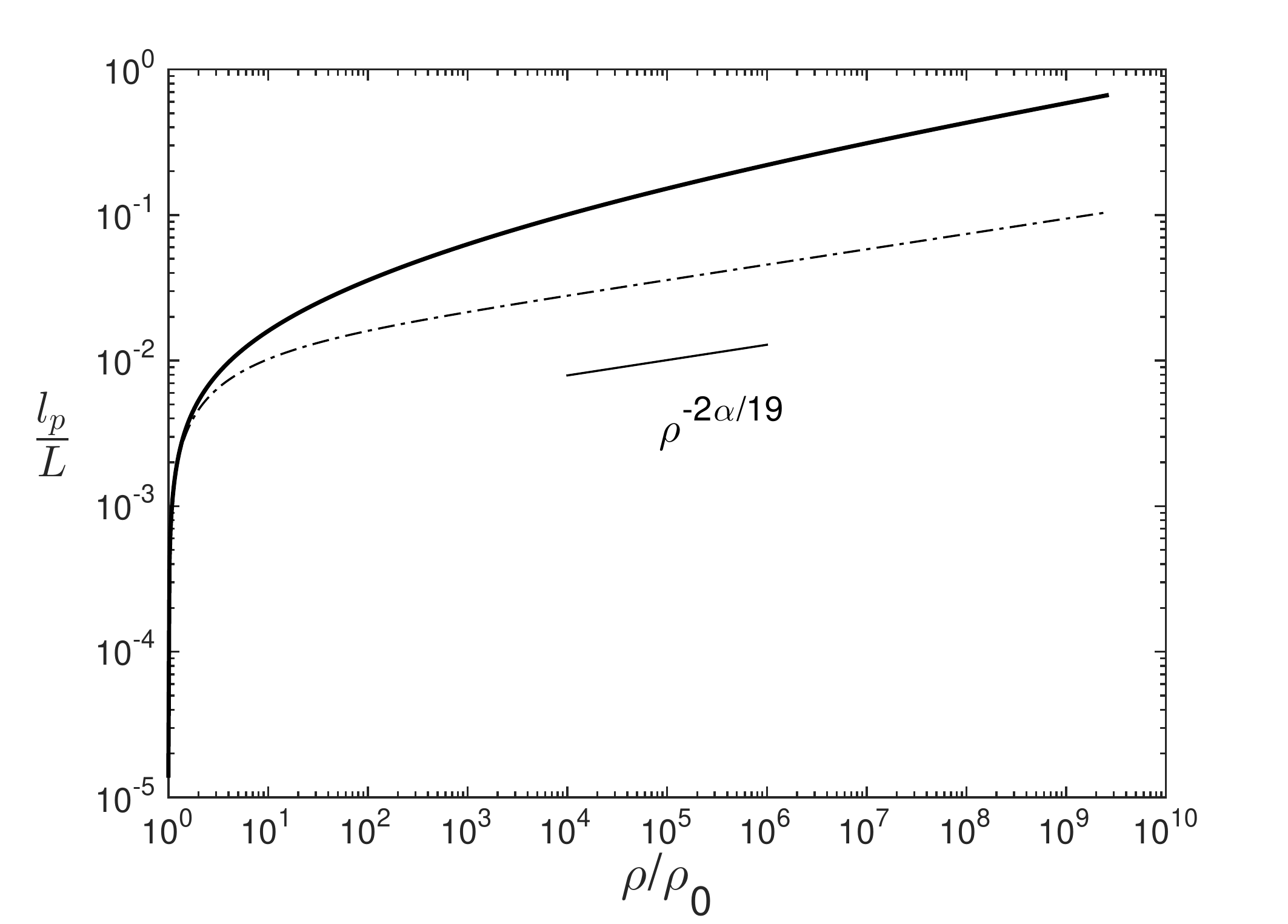}
\caption{ Normalized $l_p$ as a function of normalized $\rho$. 
The solid line is given by Eq. \eqref{eq: lpexp}.
The dash-dotted line is its approximation corresponding to the approximate expression of $\mathcal{E}$ (
the dash-dotted line in Fig. \ref{fig: ene}).
From \cite{XuLgra20}.}
\label{fig: enelp}
\end{figure}

Fig. \ref{fig: enelp} shows the evolution of correlation length $l_p = 1/k_p$ as a function of $\rho$. 
As $l_p$ is also the energy equipartition scale, it is related to $ \mathcal{E}$ by 
\begin{equation}\label{eq: maeqp}
     \mathcal{E}  = \frac{1}{2} v_p^2 = \frac{1}{2} L^{-\frac{2}{3}} V_L^2 k_p^{-\frac{2}{3}},
\end{equation}
where $v_p$ is the turbulent velocity at $k_p$, 
\begin{equation}\label{eq: kolsc}
    v_p = V_L (k_pL)^{-\frac{1}{3}}.
\end{equation}
Eq. \eqref{eq: maeqp} can be rewritten as 
\begin{equation}\label{eq: lpexp}
 \frac{l_p}{L} = \Big(\frac{\mathcal{E}}{\frac{1}{2}V_L^2}\Big)^\frac{3}{2}.
\end{equation}
With the growth of $\mathcal{E}$, 
$l_p/L$ first increases rapidly with $\rho$ and then increases gradually with the scaling close to 
(Eq. \eqref{eq: deerdw})
\begin{equation}
     \frac{l_p}{L} \propto \mathcal{E}^\frac{3}{2} \propto  \rho^{-\frac{2\alpha}{19}}
\end{equation}
up to unity.

\begin{figure*}[ht]
\centering   
   \includegraphics[width=12cm]{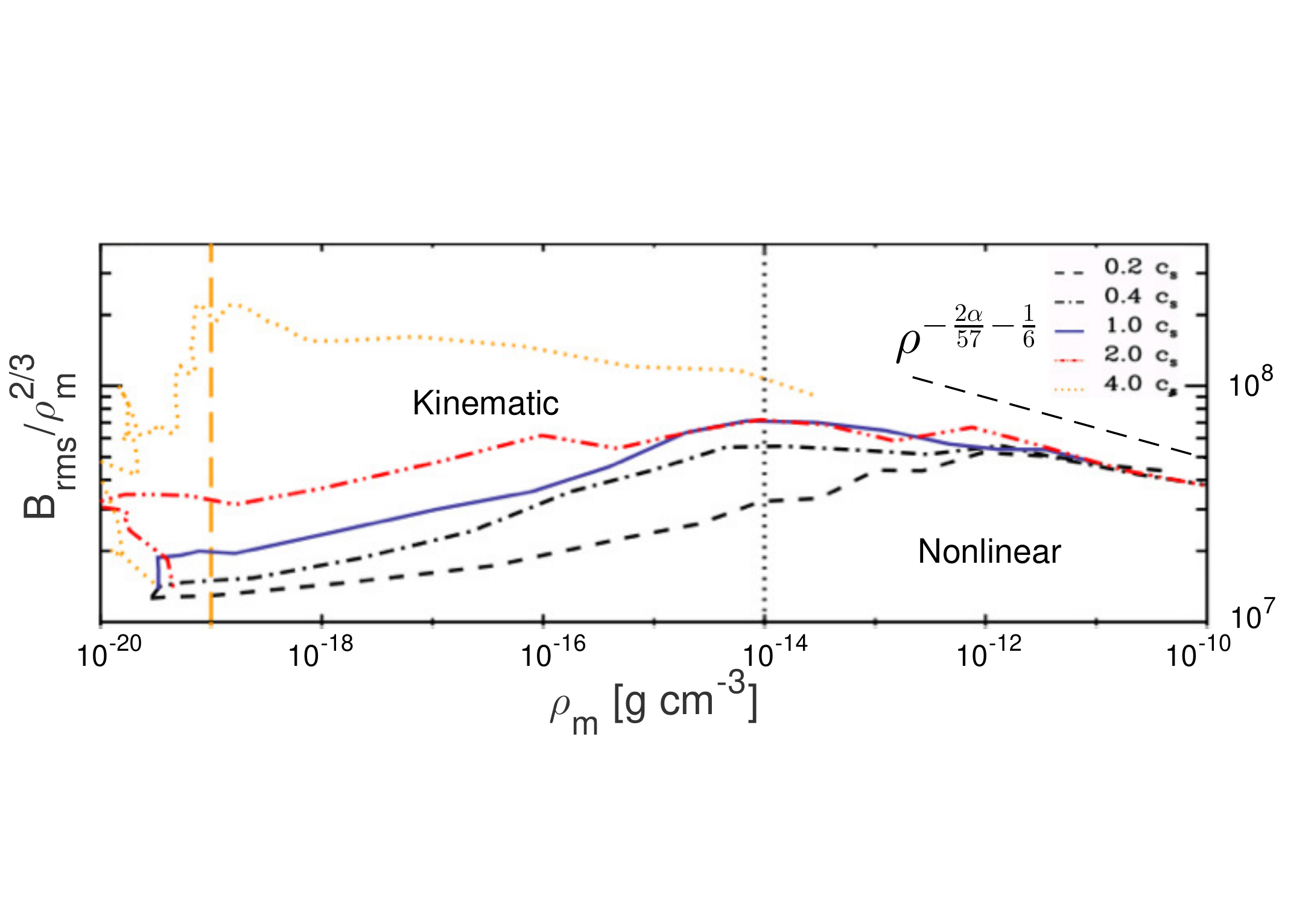}
\caption{ Comparison between the theory by \cite{XuLgra20} and the numerical result in \cite{Sur12}
on the evolution of magnetic field strength as a function of density during gravitational collapse. 
For the numerical result in \cite{Sur12}, 
different colors represent the numerical runs with different initial turbulent velocities, 
$v_{tur,0}=0.2, 0.4, 1.0, 2.0, 4.0$ $c_s$, where $c_s$ is the sound speed. 
The yellow dashed vertical line for $v_{tur,0}=4.0 c_s$ and 
the black dotted vertical line for other values of $v_{tur,0}$ indicate the transition from the kinematic to nonlinear dynamo regime. 
From \cite{XuLgra20}.}
\label{fig: fedmod}
\end{figure*}

By comparing with the 
evolution of magnetic field strength $B \propto \sqrt{ \mathcal{E}\rho}$
as a function of density numerically measured by
\cite{Sur12} from simulations on magnetic field amplification during gravitational collapse, 
the analytical expectation for nonlinear dynamo in a collapsing system (dashed line in Fig. \ref{fig: fedmod}) agrees well with the numerical 
result. 
It demonstrates the importance of 
reconnection diffusion and the breakdown of flux-freezing assumption. 
Table \ref{tab:comgrav} clearly shows the differences between the cases with reconnection diffusion and flux-freezing 
for the evolution of magnetic fields during the gravitational collapse. 
This result is important for studying the dynamics of magnetic fields and their effects on the primordial star formation
\cite{Mc20,Shar20}.

\begin{table*}[!htbp]
\renewcommand\arraystretch{1.5}
\centering
\begin{threeparttable}
\caption[]{Comparison between the cases with reconnection diffusion and flux-freezing. From \cite{XuLgra20}.}\label{tab:comgrav} 
  \begin{tabular}{c|c|c}
     \toprule
                         &            Reconnection diffusion                    &  Flux-freezing  \\
                      \hline
     \multirow{2}{*}{Magnetic energy spectrum}        &         Kazantsev spectrum ($k<k_p$)       &  \multirow{2}{*}{Kazantsev spectrum ($k<k_d$)}   \\
                                                                                &          Kolmogorov spectrum ($k_p<k<k_d $)   &   \\                                   
                      \hline
     Magnetic field structure                       &    Turbulent structure               & Folded structure   \\
                      \hline
     Correlation length                               &     \multirow{2}{*}{$1/k_p$ (equipartition scale)}      &   \multirow{2}{*}{$1/k_d$ (dissipation scale)}   \\    
     of magnetic fields & & \\          
                      \hline
  Dependence of $B$        &  $B \propto \rho^{\frac{2}{57}+\frac{1}{2}}$ (nonlinear)     &  \multirow{2}{*}{$B\propto \rho^{2/3}$}     \\
                 on $\rho$ under compression                                     &  $B\propto \rho^\frac{1}{2}$ (saturated)  &    \\
     \bottomrule
    \end{tabular}
 \end{threeparttable}
\end{table*}

We caution that for comparison with dynamo simulations in a collapsing system, 
given the limited numerical resolution, 
it is very challenging to numerically resolve the entire turbulent cascade down to the smallest turbulent eddies. 
As the fastest dynamo rate is determined by the turnover rate of the smallest eddies, 
the dynamo growth is affected and 
the kinematic dynamo in simulations is prolonged 
\cite{Mc20}
(also see Fig. \ref{fig: fedmod}).
In addition, here we adopt a Kolmogorov turbulent cascade by assuming the turbulent cascade is faster than the 
gravitational contraction. 
In the central collapsing region, when the local contraction becomes faster than the turbulent cascade,
additional turbulence can be driven with the gravitational energy converted to turbulent energy, 
which can have a different scaling from the Kolmogorov one
\cite{RobG12,Lee15,Xug20}.

\section{Turbulent dynamo in a partially ionized plasma}
\label{sec: part}

Partially ionized plasmas are very common in astrophysical environments, such as molecular clouds and 
protoplanetary disks. 
In a partially ionized medium, the coupling between magnetic fields and neutrals is determined by the coupling between neutrals and ions
via their collisions. 
Table \ref{Tab: dam} lists different coupling regimes for both MHD turbulence with strong magnetic fields and 
turbulent dynamo with initially weak magnetic fields. 
Here $\nu_{in}$ and $\nu_{ni}$ are ion-neutral and neutral-ion collisional frequencies. 
In a weakly ionized medium, there is $\nu_{in} > \nu_{ni}$. 
$\omega_A$ is the Alfv$\acute{e}$n wave frequency, 
and $\Gamma_l$ is the dynamo stretching rate at length scale $l$. 
When the neutral-ion collisions are infrequent, the relative drift between the two species causes diffusion of magnetic fields, i.e., ambipolar diffusion,
and inefficient dynamo growth of magnetic fields. 
In a weakly ionized medium, the 
ion-neutral collisional damping is also the dominant damping effect for magnetic fluctuations 
\cite{XLY14,Xuc16,XL17}.

\begin{table}[ht]
\renewcommand\arraystretch{1.3}
\centering
\begin{threeparttable}
\caption[]{ Coupling regimes in MHD turbulence and turbulent dynamo. From \cite{Xud19}.
}\label{Tab: dam} 
  \begin{tabular}{cccc}
      \toprule
    Coupling regime                            &    Strong coupling                  &    Weak coupling                                    & Decoupling     \\
                                                              \midrule
     MHD turbulence       &     $ \omega_A < \nu_{ni}$     &   $ \nu_{ni} < \omega_A < \nu_{in} $     &  $\omega_A > \nu_{in}$   \\
             Turbulent dynamo               &     $\Gamma_l < \nu_{ni}$        &    $\nu_{ni} < \Gamma_l < \nu_{in} $       &  $\Gamma_l > \nu_{in}$   \\
 \bottomrule
    \end{tabular}
 \end{threeparttable}
\end{table}

\subsection{Turbulent dynamo in a weakly ionized medium}

XL16 introduced a parameter 
\begin{equation}\label{eq: ratgam}
  \mathcal{R} 
  = \frac{6}{\xi_n} \frac{\nu_{ni}}{\Gamma_\nu},
\end{equation}
where $\xi_n$ is the neutral fraction, $\nu_{ni}$ is the neutral-ion collisional frequency, 
and $\Gamma_\nu$ is the dynamo stretching rate at the viscous scale. 
$\mathcal{R}$ depends on the ionization fraction and reflects the strength of neutral-ion coupling. 
At $\mathcal{R} > 1$ with a high ionization fraction, the neutral-ion coupling is sufficiently strong for the turbulent motions 
carried by neutrals at the viscous scale to efficiently stretch and amplify magnetic fields. 
At $\mathcal{R} < 1$ with a low ionization fraction, the neutral-ion coupling is weak, which results in ineffective turbulent 
stretching and inefficient dynamo growth of magnetic fields.

XL16 found that the turbulent dynamo at $\mathcal{R} < 1$ is characterized by a unique damping stage of dynamo. 
During the damping stage, magnetic energy grows as 
\begin{equation}\label{eq: emdt}
   \sqrt{ \mathcal{E}_M} = \sqrt{ \mathcal{E}_{M1}} + \frac{3}{23} \mathcal{C}^{-\frac{1}{2}} L^{-\frac{1}{2}} 
   V_L^\frac{3}{2} (t - t_1) ,
\end{equation}
where $\mathcal{E}_{M1}$ is the magnetic energy at the beginning of the damping stage at $t = t_1$, and 
$\mathcal{C}= \xi_n / (3 \nu_{ni})$ is related to the ionization fraction. 
It means that the magnetic field strength $B \propto \sqrt{ \mathcal{E}_M}$ grows linearly with time, and the growth rate 
depends on the ionization fraction. 
We note that due to the weak coupling between neutrals and ions, the growth of $\mathcal{E}_M$ mainly comes from the turbulent energy contained in ions.

\begin{figure}[ht]
\centering   
   \includegraphics[width=8.5cm]{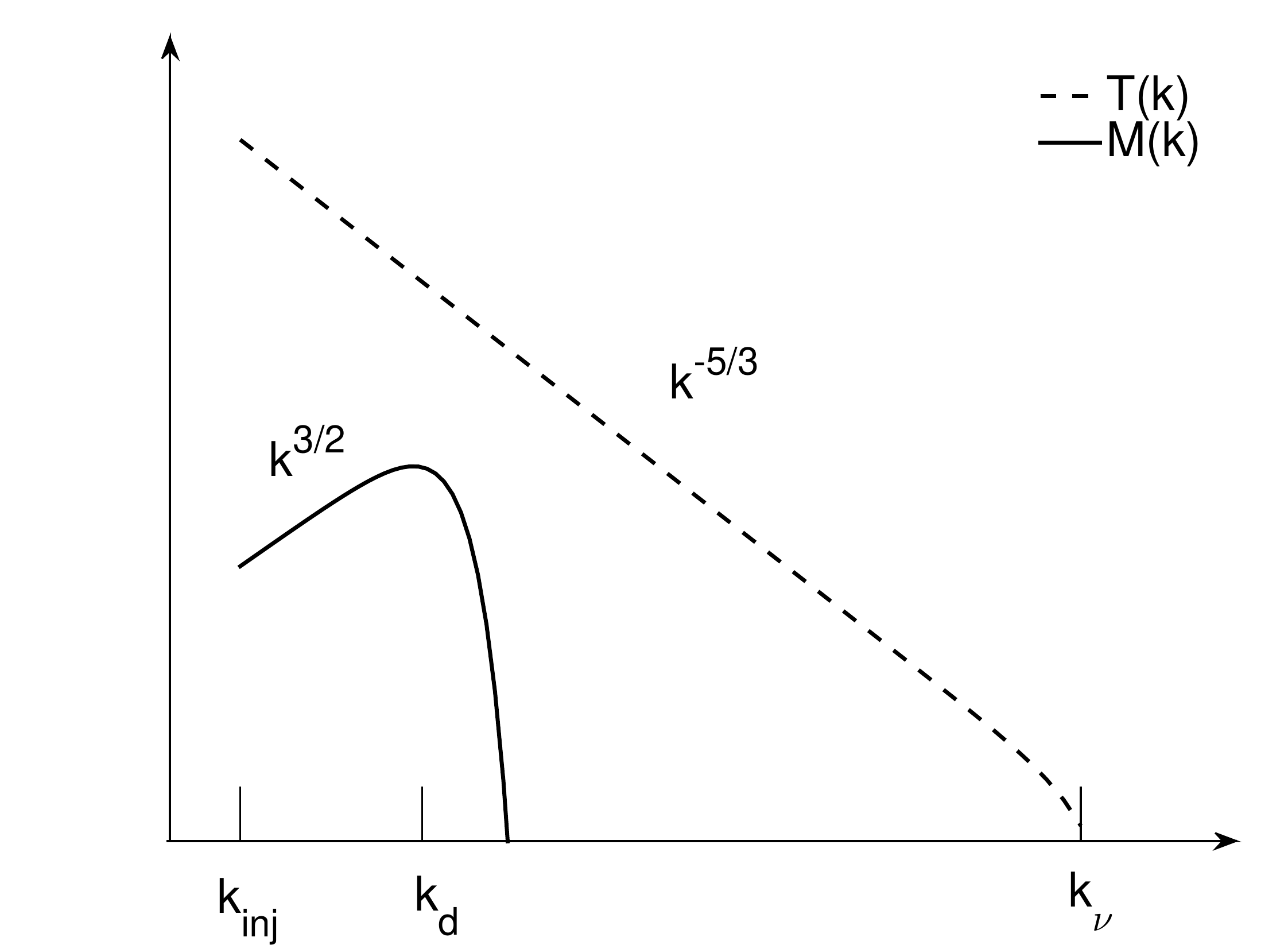}
\caption{Illustration for the magnetic energy spectrum $M(k)$ and the turbulent energy spectrum $T(k)$ in the  
damping stage of turbulent dynamo. From \cite{Xud19}. }
\label{fig: skd}
\end{figure}

As shown in Fig. \ref{fig: skd}, the magnetic energy spectrum during the damping stage is cut off at the ion-neutral collisional damping scale $l_d = 1/k_d$, 
where significant damping of magnetic fluctuations takes place. 
With the growth of magnetic energy, $l_d$ evolves as 
\begin{equation}\label{eq: ldevlt}
    l_d = \Big(l_{d1}^\frac{2}{3} + \frac{3}{23} L^{-\frac{1}{3}} V_L (t-t_1)\Big)^\frac{3}{2},
\end{equation}
where $l_{d1}$ is the damping scale at $t=t_1$.
As $l_d$ is also the correlation length of the amplified magnetic fields. It means that the magnetic fields have an increasing correlation length 
during the damping stage.

\subsection{Numerical test with two-fluid dynamo simulation}

To numerically test the turbulent dynamo theory in a weakly ionized medium, it is necessary to carry out 
a two-fluid simulation that treats the dynamics of neutrals and ions separately. 
\cite{Xud19}
performed a 3D high-resolution two-fluid dynamo simulation to 
solve the following equations
\cite{Dra86}
by using the two-fluid RIEMANN code 
\cite{Bal98a,Bal98b},
\begin{equation}
\begin{aligned}
&\frac{\partial \rho_i}{\partial t} + \nabla \cdot (\rho_i \boldsymbol{v}_i) = 0,  \\
&\frac{\partial \boldsymbol{v}_i}{\partial t} + (\boldsymbol{v}_i \cdot \nabla)\boldsymbol{v}_i =
     -c_s^2 \nabla \ln \rho_i - \frac{1}{4\pi}\boldsymbol{B}\times(\nabla \times \boldsymbol{B})  \\
&  ~~~~~~~~~~~~~~~~~~~~~~~~~~~~~~~~   - \gamma_d \rho_n (\boldsymbol{v}_i - \boldsymbol{v}_n),  \\
&\frac{\partial \boldsymbol{B}}{\partial t} = \nabla \times ( \boldsymbol{v}_i \times \boldsymbol{B}), \\
&\frac{\partial \rho_n}{\partial t} + \nabla \cdot (\rho_n \boldsymbol{v}_n) = 0,  \\
&\frac{\partial \boldsymbol{v}_n}{\partial t} + (\boldsymbol{v}_n \cdot \nabla)\boldsymbol{v}_v =
      -c_s^2 \nabla \ln \rho_n - \gamma_d \rho_i (\boldsymbol{v}_n - \boldsymbol{v}_i), 
\end{aligned}
\end{equation}
where $\boldsymbol{v_i}$ and $\boldsymbol{v_n}$ are the velocities of the ionized and neutral fluids, 
$\boldsymbol{B}$ is the magnetic field,
$c_s$ is the sound speed,
$\rho_i$ and $\rho_n$ are ion and neutral mass densities, 
and $\gamma_d$ is the drag coefficient 
\cite{Shu92}.
Table \ref{tab: par} lists the parameters used in the simulation, where $R$ is the numerical resolution, and
$V_\text{rms}$ is the injected turbulent velocity at the driving scale $L$.
The initial large Alfv$\acute{e}$n Mach number $M_{Ai0}$ corresponds to the initial weak magnetic field for a dynamo simulation. 
The ionization fraction should be sufficiently small to ensure that neutrals are weakly coupled with ions and 
the dynamo is in the damping regime. 

\begin{table}[t]
\renewcommand\arraystretch{1.5}
\centering
\begin{threeparttable}
\caption[]{Simulation parameters}\label{tab: par} 
  \begin{tabular}{cccccc}
     \toprule
       $R$                &    $L$                   &    $\rho_i/\rho_n$              & $V_\text{rms}$  & $c_s$  &    $M_{Ai0}$         \\
       $1024^3$       &    $512$               &     $1.26\times10^{-3}$      &  $0.2$               &  $1$     &     $17.7$                \\
    \bottomrule
    \end{tabular}
 \end{threeparttable}
\end{table}

\begin{figure}[ht]
\centering    
   \includegraphics[width=8.5cm]{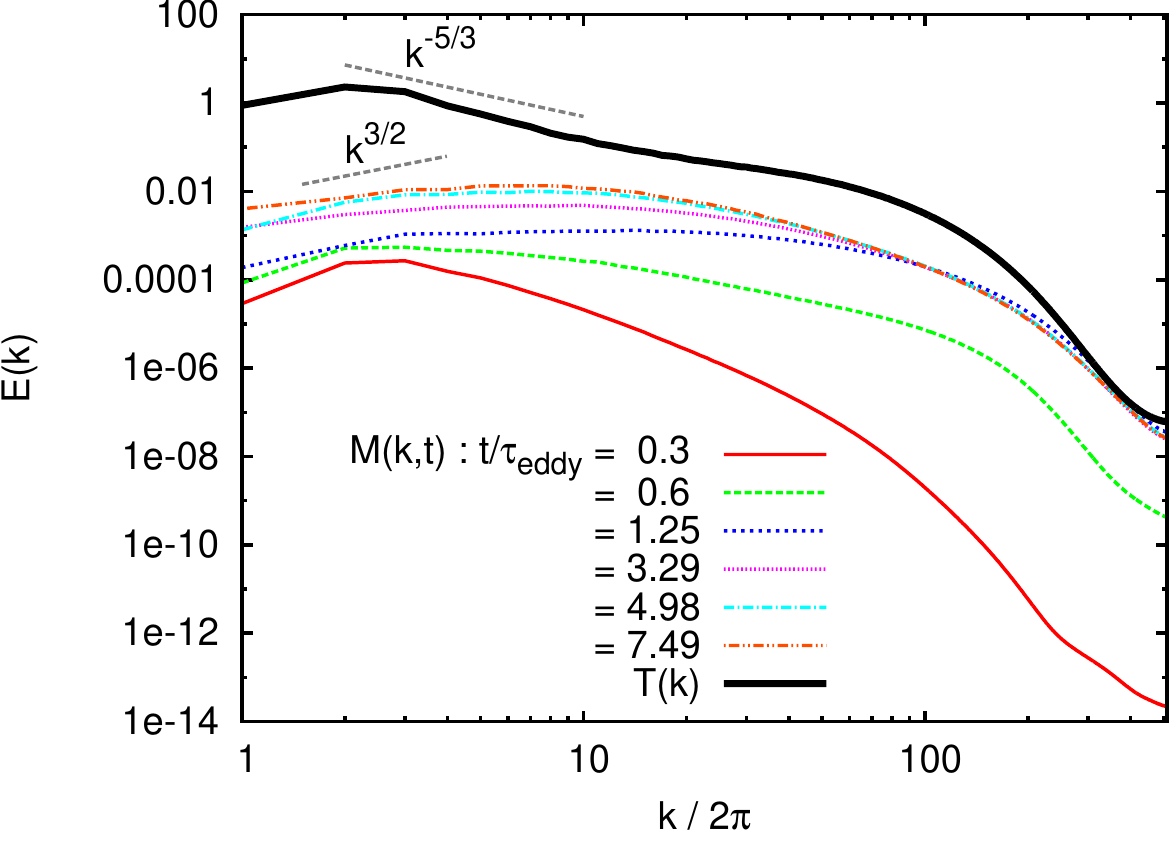}
\caption{Numerically measured $M(k,t)$ at different times during the two-fluid dynamo simulation. 
$T(k)$ is the energy spectrum of driven turbulence. From \cite{Xud19}.   }
\label{fig: numev}
\end{figure}

\begin{figure*}[ht]
\centering   
   \subfigure[]{
   \includegraphics[width=7.5cm]{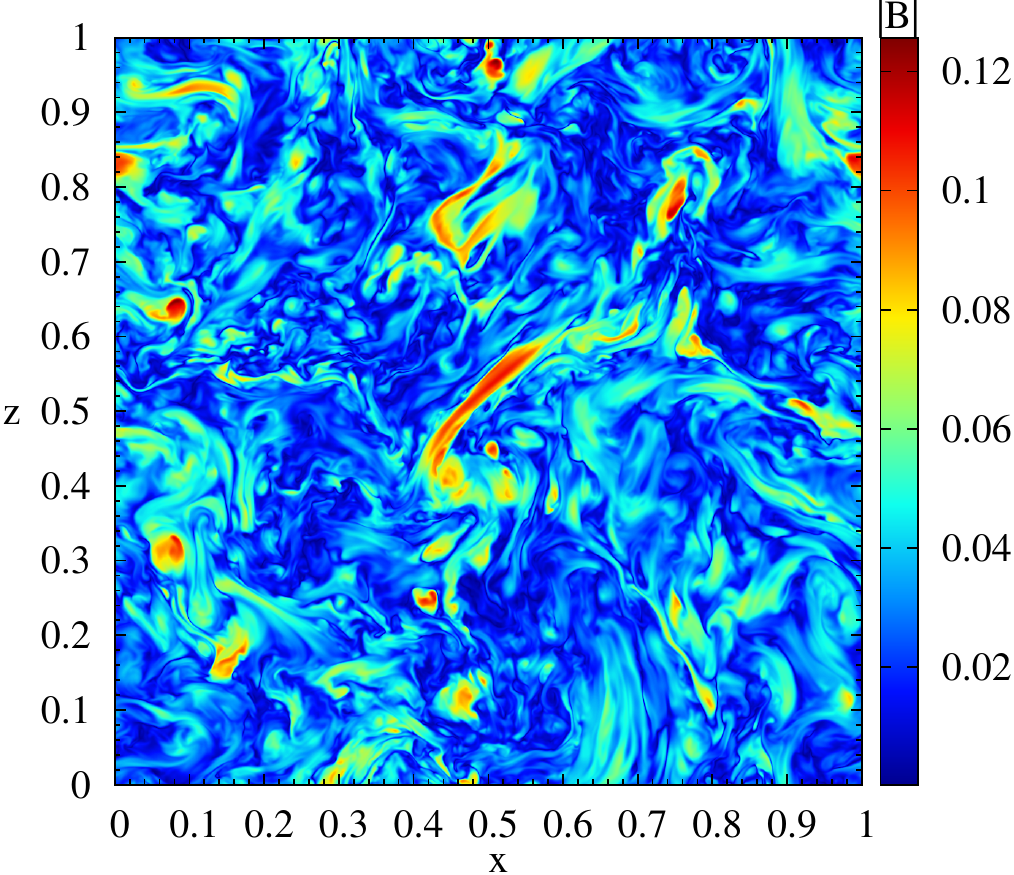}\label{fig: magstra}}
   \subfigure[]{
   \includegraphics[width=7.5cm]{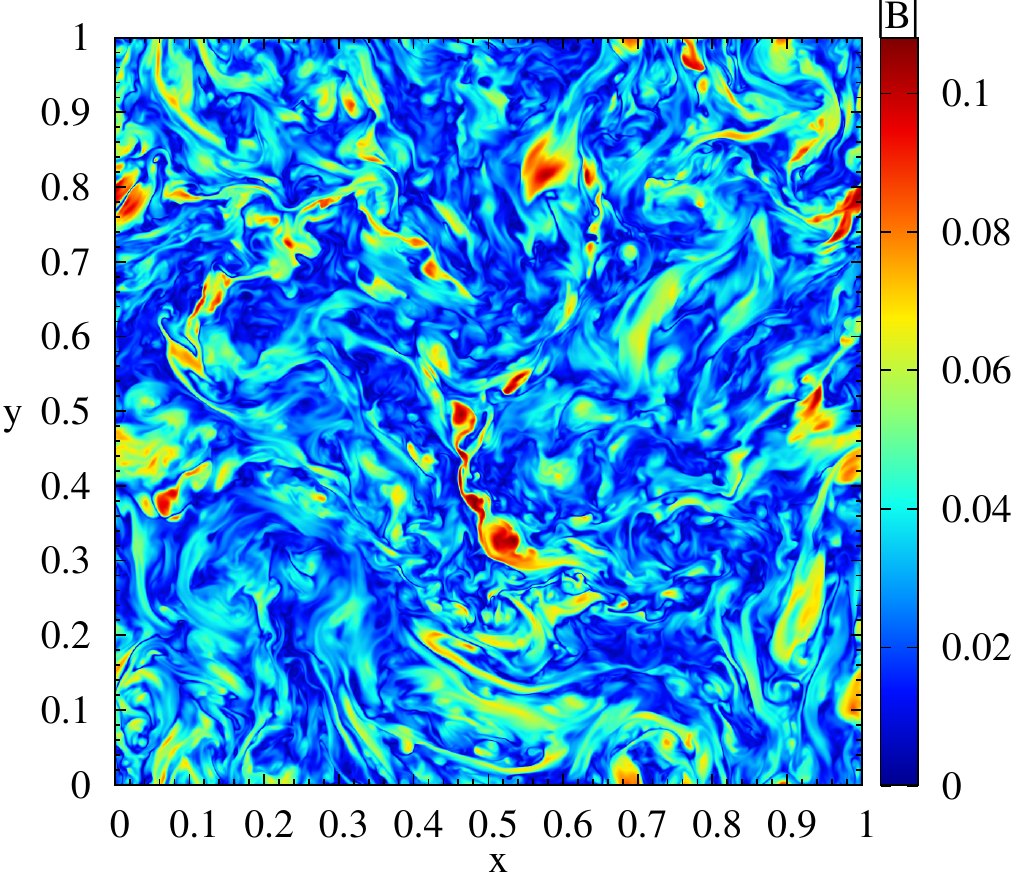}\label{fig: magstrb}}
   \subfigure[]{
   \includegraphics[width=7.5cm]{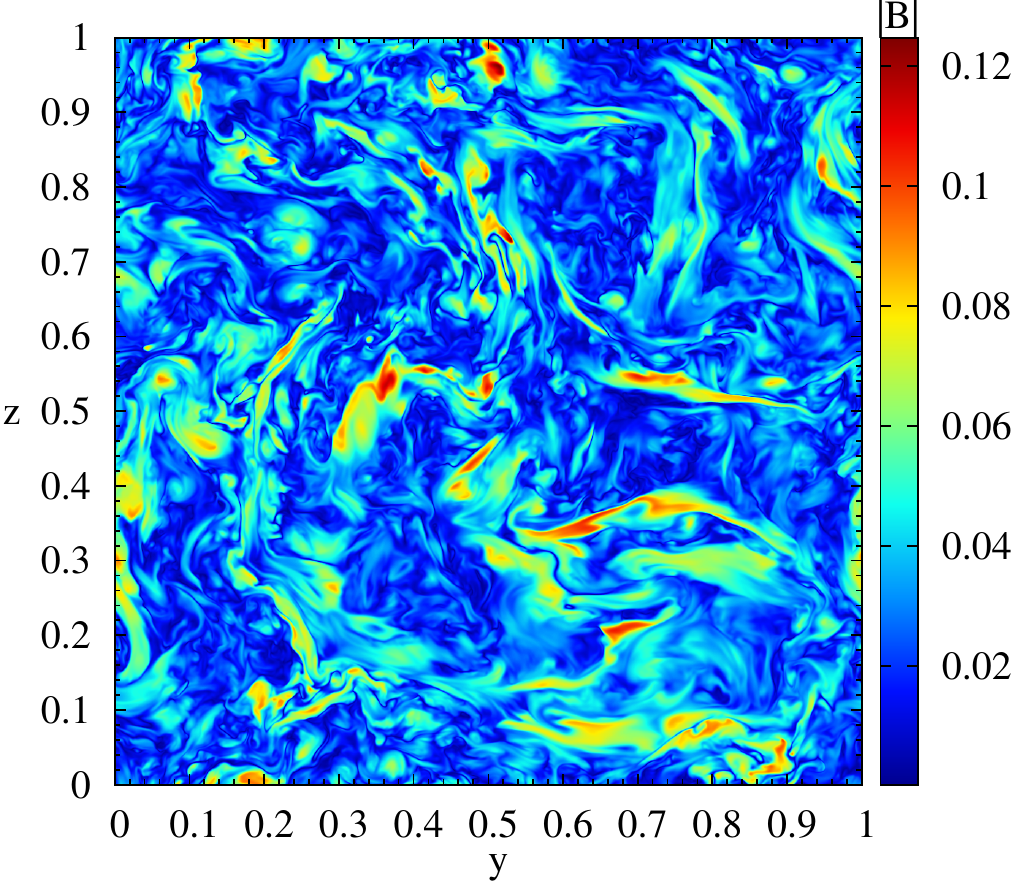}\label{fig: magstrc}}
\caption{2D cross sections of the numerically measured magnetic field strength
in the (a) xz plane, (b) xy plane, and (c) yz plane at $t=7.49~\tau_\text{eddy}$.
From \cite{Xud19}. }
\label{eq: ms2dldst}
\end{figure*}

Fig. \ref{fig: numev} shows the numerically measured magnetic energy spectrum at different times of the simulation. 
The initial uniform magnetic field is aligned along the x-direction. 
The smallest turbulent eddies have the fastest dynamo stretching rate. 
So the peak scale of the initial magnetic energy spectrum $M(k,t)$ first shifts to smaller scales. 
Then with the growth of magnetic energy, the ion-neutral collisional damping effect becomes more significant. 
Accordingly, the damping scale $l_d$, where $M(k,t)$ peaks, 
increases with time (see Eq. \eqref{eq: ldevlt}). 
At $t=7.49~\tau_\text{eddy}$, where $\tau_\text{eddy}$ is the eddy-turnover time at $L$, 
the amplified magnetic fields have a large correlation length. 
The smaller-scale magnetic fluctuations on scales less than $l_d$ are suppressed due to ion-neutral collisional damping. 
This can be clearly seen from the magnetic field structure shown in Fig. \ref{eq: ms2dldst}.

\begin{figure}[htbp]
\centering   
   \includegraphics[width=8.5cm]{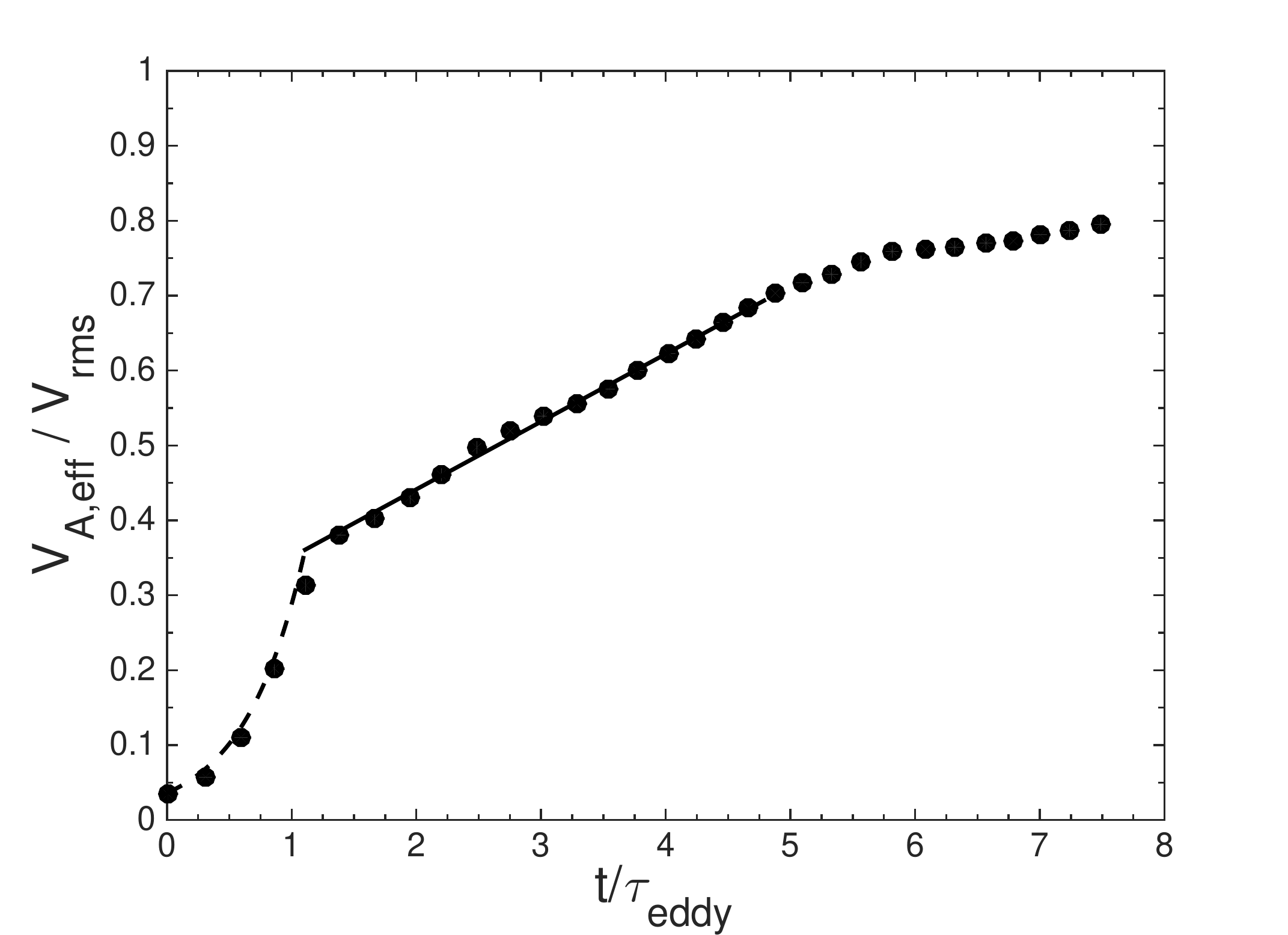}
\caption{  Time evolution of the numerically measured $V_{A,\text{eff}}$ (normalized by $V_\text{rms}$, circles) 
in comparison with the theoretical prediction (solid line). 
From \cite{Xud19}.}
\label{fig: ed}
\end{figure}

As an important characteristic of the turbulent dynamo in a weakly ionized medium, 
the analytically predicted linear-in-time growth of magnetic field strength $B$ during the damping stage of dynamo (see Eq. \eqref{eq: emdt})
is also numerically confirmed in \cite{Xud19}. 
As displayed in Fig. \ref{fig: ed}, after the exponential growth of $B$ during the first $\tau_\text{eddy}$, 
$B$ grows linearly with time in the subsequent damping stage. 
The effective Alfv$\acute{e}$n speed $V_{A,\text{eff}}$ defined in \cite{Xud19}
is proportional to $B$.

The above numerically tested dynamo theory in a weakly ionized medium can be generally applied to studying magnetic field amplification 
in weakly ionized astrophysical media. 
The dynamo growth of magnetic fields and their properties 
depend on the turbulence parameters and ionization fraction in the local environment.

\subsection{Magnetic field amplification in supernova remnants: preshock region}

Supernova shocks are believed to be the accelerators of Galactic CRs
\cite{Bla87}, 
and magnetic fields much stronger than 
the interstellar magnetic fields are required for the particle confinement and acceleration (see Section \ref{ssec: shpost}). 
Magnetic field amplification in the preshock region in the literature is usually attributed to the CR-driven instabilities 
\cite{Bell2004}.
More recently, turbulent dynamo has also been invoked as a different mechanism for magnetic field amplification in the 
preshock region
\cite{BJL09,Dru12,Del16}.
The upstream turbulence can be induced by the interaction between the 
CR pressure gradient in the shock precursor and the upstream density inhomogeneities
\cite{BJL09,Dru12}.
Density inhomogeneities
are very common in the interstellar medium as suggested by observations 
\cite{XuZ17,XZ20}.
\cite{XuL17}
took into account the effect of ion-neutral collisional damping in a partially ionized medium 
and reexamined the turbulent dynamo in the shock precursor.

\begin{figure}[htbp]
\centering   
   \includegraphics[width=8.5cm]{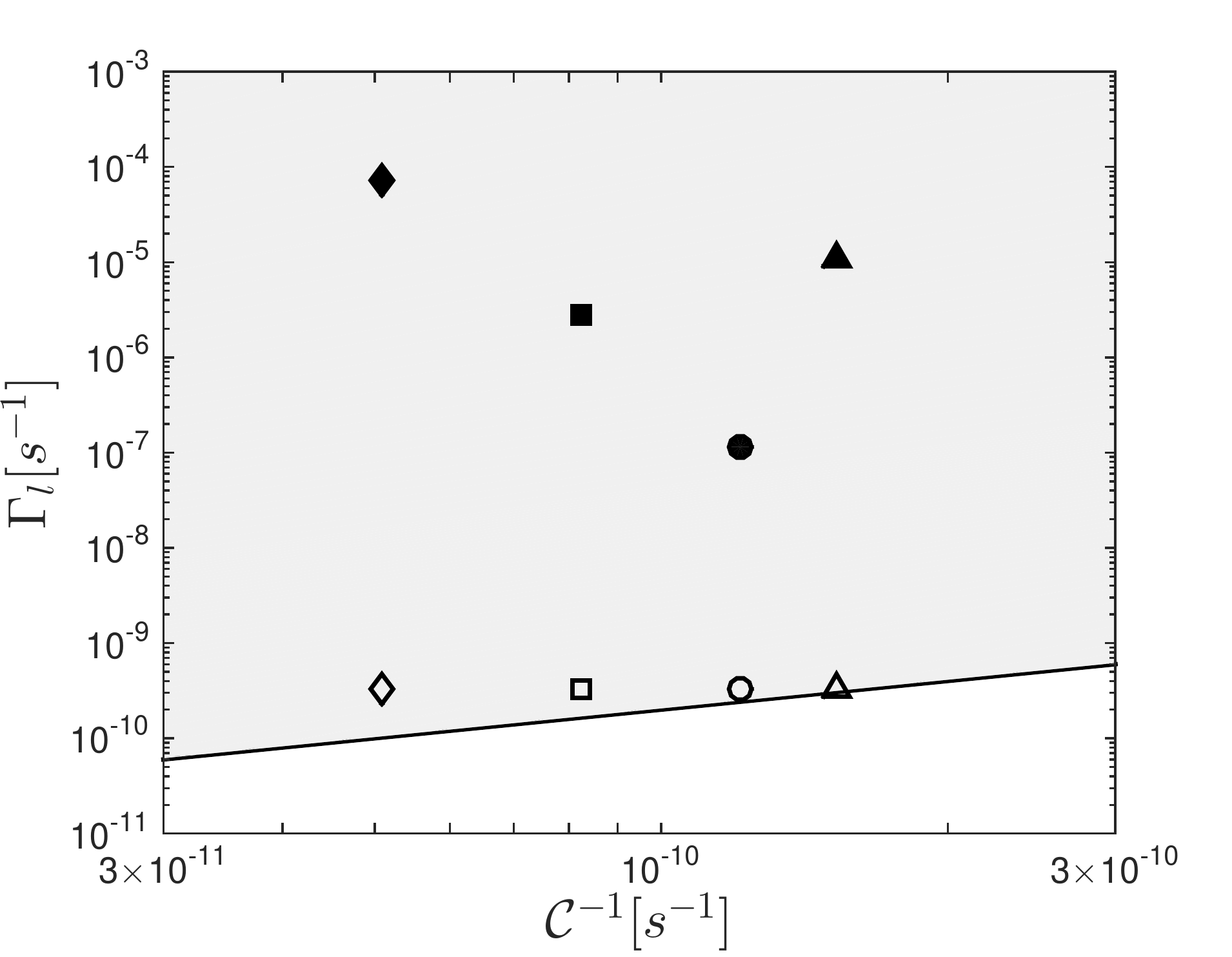}
\caption{Dynamo stretching rate $\Gamma$ vs. $\mathcal{C}^{-1}$ for the turbulent dynamo in the shock precursor
in the WNM (circle), CNM (square), MC (triangle), and DC (diamond). 
The shaded region shows the 
parameter space for the damping stage of dynamo. 
Filled and open symbols represent $\Gamma$ at the viscous scale and $L$, respectively.  
From \cite{Xud19}. }
\label{fig: ismts}
\end{figure}

For supernova shocks propagating in the multi-phase interstellar medium, including 
the warm neutral medium (WNM), cold neutral medium (CNM), MCs, and dense cores (DC) (Table \ref{Tab: ism}), 
by considering the driving condition of the precursor turbulence as 
\begin{equation}
       L=0.1\, \text{pc},   ~V_L=10^3 \,\text{km s}^{-1},
\end{equation}
where $L$ is the characteristic scale of the density structure in cold interstellar phases
\cite{Hei03,go98} 
and $V_L$ is of the order of shock velocity, 
\cite{Xud19}
found that the turbulent dynamo in the shock precursor is entirely in the damping stage as shown in 
Fig. \ref{fig: ismts}.
Here $n_H$ and $n_e$ are number densities of the atomic hydrogen and electrons, and $T$ is the temperature
\cite{DraL98}.
The timescale of the dynamo $\tau_\text{dam}$, 
\begin{equation}
    \tau_\text{dam} = \frac{23}{3} \Big(\frac{L}{V_L} - \Gamma_\nu^{-1}   \Big) ,
\end{equation}
where $\Gamma_\nu$ is the dynamo stretching rate at the viscous scale, 
and the strength $B_\text{dam}$ and correlation length $l_{d,\text{cr}}$ of magnetic fields 
reached at the end of dynamo are all listed in Table \ref{Tab: ism}. 
Magnetic fields of the order of $100~ \mu$G near the shock front are indicated by observations 
(e.g., \cite{Bam03,Bam05,Bamb05,Vin12}).

As $\tau_\text{dam}$
is small compared with the precursor crossing time $(c/v_\text{sh}) L/V_L$, where $c$ and $v_\text{sh}$ are the
light speed and shock propagation speed, 
the dynamo-amplified magnetic fields with the correlation length $L$ can be present in the shock precursor.

The diffusion of CRs is significantly affected by the structure of magnetic fields. 
For CRs streaming along the dynamo-amplified magnetic fields in a weakly ionized medium, 
their effective mean free path is determined by the ion-neutral collisional 
damping scale. 
The maximum energy of CRs that can be confined in the case of the shock propagation in an MC is 
\begin{equation}\label{eq: maxcrene}
    E_\text{CR,max} = e B_\text{dam} L \approx 38 ~\text{PeV}.
\end{equation}
Different from the magnetic field amplification induced by CR-driven instabilities, 
the magnetic fields amplified by turbulent dynamo 
have the correlation length independent of CR gyroradius. 
Both mechanisms should be taken into account when studying CR diffusion and acceleration at supernova shocks.

For supernova shocks propagating through the partially ionized interstellar 
medium, 
we discuss the turbulent dynamo as a mechanism for amplifying the preshock magnetic fields. 
Given the parameters adopted here, Eq. \eqref{eq: maxcrene} provides an estimate for the maximum energy of CRs 
that the dynamo-amplified magnetic fields can confine. 
The detailed modeling of shock acceleration of CRs that involves the dynamo-amplified magnetic fields 
and the dependence of the maximum CR energy on the local interstellar environment
require further studies.
We also note that the recent observations of SNRs suggest that the maximum energy of accelerated CRs is below 1 PeV
(e.g., \cite{Ahn17,Abda18}). 
In these observed SNRs, 
the magnetic field amplification may not be very efficient.

\begin{table}[h]
\renewcommand\arraystretch{1.3}
\centering
\begin{threeparttable}
\caption[]{Turbulent dynamo in the shock precursor in the multi-phase interstellar medium. From \cite{Xud19}.
}\label{Tab: ism} 
  \begin{tabular}{ccccc}
      \toprule
 & WNM & CNM & MC & DC \\
      \midrule
$n_\text{H}$[cm$^{-3}$]  & $0.4$ & $30$ & $300$ & $10^4$  \\
$n_e/n_\text{H}$  & $0.1$ & $10^{-3}$ & $10^{-4}$ & $10^{-6}$  \\
$T$[K]  & $6000$ & $100$ & $20$ & $10$  \\
\cline{1-5}
$l_{d,\text{cr}}$ [pc]  & \multicolumn{4}{c}{$0.1$}   \\
$\tau_\text{dam}$[kyr]   &   \multicolumn{4}{c}{$0.75$}  \\
$B_\text{dam}$ [$\mu$ G]  & $79.1$  & $56.6$  & $415.2$  & $138.2$ \\
 \bottomrule
    \end{tabular}
 \end{threeparttable}
\end{table}

\subsection{Multiple evolutionary stages of turbulent dynamo in a partially ionized medium }

Besides the damping stage of dynamo, depending on the ionization fraction and initial magnetic field strength, 
the turbulent dynamo in a partially ionized medium can have multiple evolutionary stages with distinctive growth behavior of 
magnetic fields. 
Tables \ref{tab: reg1}-\ref{tab: reg4} present the dependence of magnetic energy $\mathcal{E}$ on time $t$ in 
different evolutionary stages of turbulent dynamo in a partially ionized medium 
corresponding to different ranges of $\mathcal{R}$ (Eq. \eqref{eq: ratgam}). 
$\mathcal{E}_0$ is the initial magnetic energy, and 
$E_{k,\nu}$ is the turbulent energy at the viscous scale. 
The detailed derivation and analysis of each evolutionary stage can be found in XL16.

\begin{table}[t]
\renewcommand\arraystretch{1.7}
\centering
\begin{threeparttable}
\caption[]{$\mathcal{R}<1$. From \cite{XL17}.}\label{tab: reg1} 
  \begin{tabular}{c|c|c|c|c}
      \toprule
        Stages                      &   Dissipation-free  &     Viscous    &    Damping      &   Nonlinear           \\
     \hline
   $\mathcal{E}$
    &  $\sim e^{2\Gamma_\nu t}$   & $\sim e^{\frac{1}{3}\Gamma_\nu t}$   & $\sim t^2$                          & $\sim t$                               \\
    
    \bottomrule
    \end{tabular}
 \end{threeparttable}
\end{table}

\begin{table}[t]
\renewcommand\arraystretch{1.7}
\centering
\begin{threeparttable}
\caption[]{$1<\mathcal{R}<5^{\frac{4}{5}} \Big(\frac{E_{k,\nu}}{\mathcal{E}_0}\Big)^\frac{1}{2}$. From \cite{XL17}. }\label{tab: reg3} 
  \begin{tabular}{c|c|c|c|c}
      \toprule
     Stages                      &   Dissipation-free  &     Viscous   &     Transitional  &  Nonlinear           \\
    \hline
    $\mathcal{E}$
    &  $\sim e^{2\Gamma_\nu t}$ &  $\sim e^{\frac{1}{3}\Gamma_\nu t}$ &  $E_{k,\nu}$   &  $\sim t$         \\
    \bottomrule
    \end{tabular}
 \end{threeparttable}
\end{table}

\begin{table}[t]
\renewcommand\arraystretch{1.7}
\centering
\begin{threeparttable}
\caption[]{$\mathcal{R}\geq 5^{\frac{4}{5}} \Big(\frac{E_{k,\nu}}{\mathcal{E}_0}\Big)^\frac{1}{2}$. From \cite{XL17}.}\label{tab: reg4} 
  \begin{tabular}{c|c|c|c}
      \toprule
     Stages                      &   Dissipation-free &  Transitional   &  Nonlinear           \\
    \hline
     $\mathcal{E}$
    &  $\sim e^{2\Gamma_\nu t}$  &  $E_{k,\nu}$ &  $\sim t$                        \\
       \bottomrule
    \end{tabular}
 \end{threeparttable}
\end{table}

\begin{figure*}[htbp]
\centering
\subfigure[]{
   \includegraphics[width=8cm]{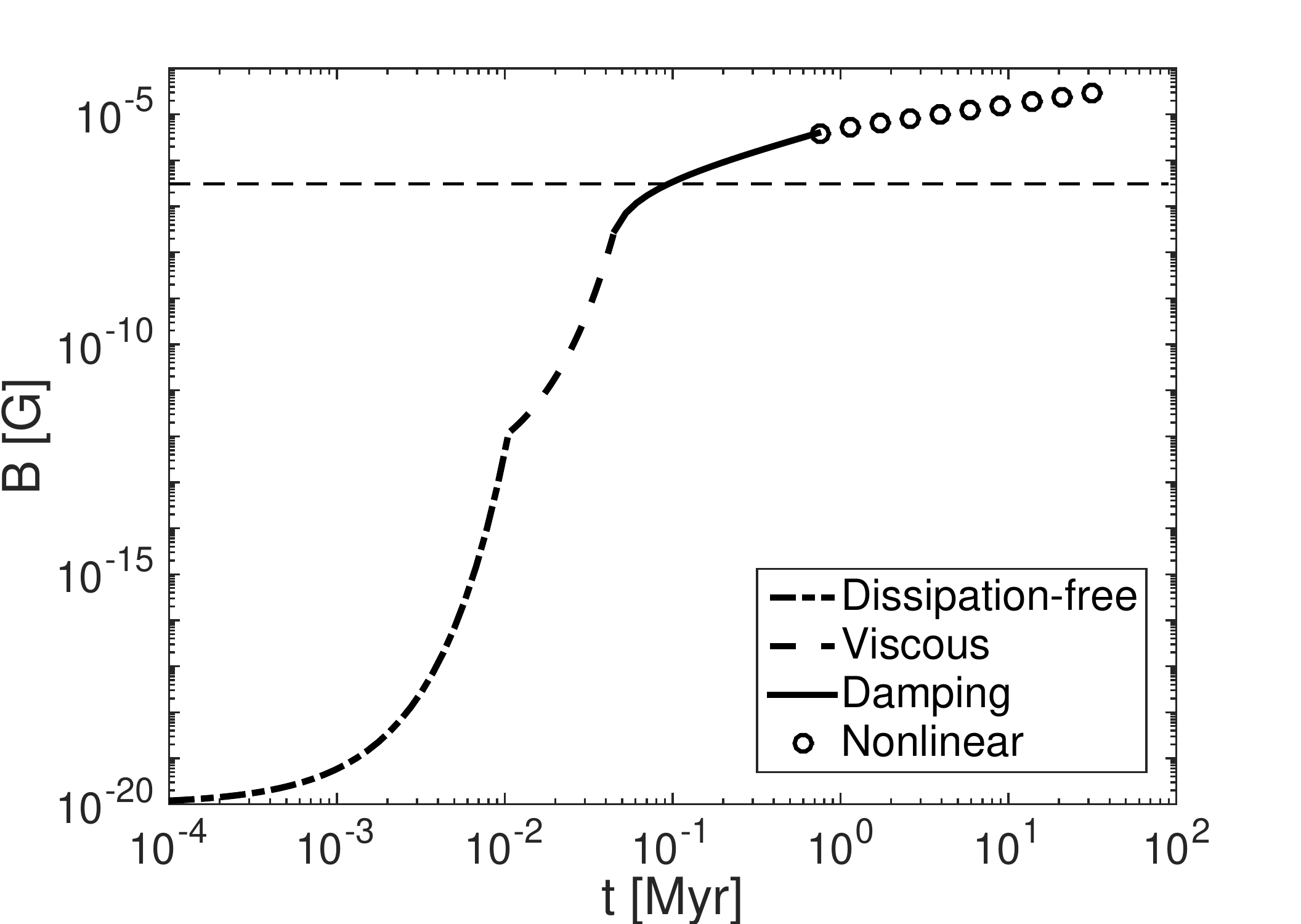}\label{fig: fgb}}
\subfigure[]{
   \includegraphics[width=8cm]{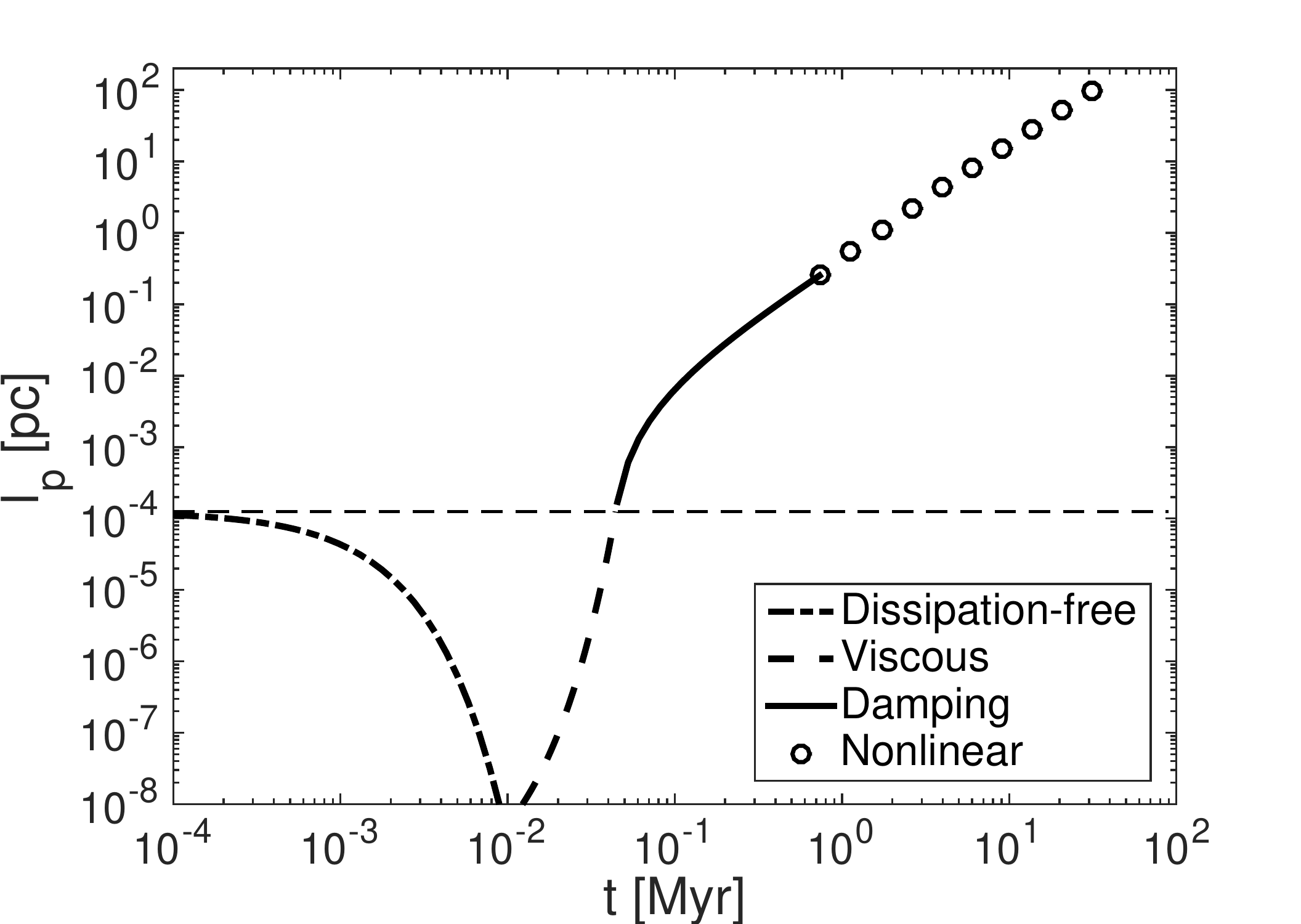}\label{fig: fgs}}  
\caption{Time evolution of the magnetic field strength and the correlation length of the amplified magnetic fields 
during the formation of the 
first galaxies. From XL16.}
\label{fig: firs}
\end{figure*}

As an illustration, XL16 considered the magnetic field amplification in the first galaxies. 
The parameters for the turbulence driving and physical conditions are taken from  
\cite{SchoSch12,Schob13}. 
The dynamo amplification of magnetic field strength $B$ and the evolution of the correlation length $l_p$ of magnetic fields are displayed in 
Fig. \ref{fig: firs}. 
Their values in corresponding evolutionary stages are provided in Table \ref{tab: fga}. 
$t_\text{ff}$ is the free-fall time of the system. 
The growth of $B$ is inefficient in both damping stage and nonlinear stage, but $l_p$ increases from the viscous scale up to the 
driving scale of turbulence during these two stages. 
The dashed lines in Fig. \ref{fig: fgb} and Fig. \ref{fig: fgs} indicate the 
magnetic field strength when the magnetic energy is in equipartition with the turbulent energy at the viscous scale 
and the viscous scale, respectively.

As the new turbulent dynamo theory was adopted, different from earlier studies by, e.g., \cite{SchoSch12,Schob13}, 
XL16 identified multiple evolutionary stages including the new damping stage of dynamo. 
For the nonlinear dynamo, its inefficiency due to reconnection diffusion of magnetic fields 
leads to a much more prolonged evolutionary stage compared with earlier studies (see Section \ref{sec: noda}). 
We note that the timescale of nonlinear stage of dynamo can be even longer than $t_\text{ff}$. 
This result can have important implications for the effect of 
magnetic fields during the formation and evolution of the first galaxies.

\begin{table*}[t]
\renewcommand\arraystretch{1.5}
\centering
\begin{threeparttable}
\caption[]{$t_\text{ff}=16.3$ Myr,  $\mathcal{R}=0.006$. From XL16.}\label{tab: fga} 
  \begin{tabular}{c|c|c|c}
      \toprule
    $B(t)$                & $t$ [Myr]  & $l_p$ [pc]   & $B$ [G]    \\
                    \hline
      Dissipation-free ($\sim e^{\Gamma_\nu t}$)
 & $1.1\times 10^{-2}$  & $6.0\times10^{-9}$   & $1.2\times10^{-12}$  \\
 \hline
Viscous ($\sim e^{\frac{1}{6}\Gamma_\nu t}$)
 &  $4.4\times10^{-2}$  & $1.2\times10^{-4}$   & $2.5\times10^{-8}$  \\
 \hline
Damping ($\sim t$)
 & $7.3\times10^{-1}$  & $2.5\times10^{-1}$  & $3.9\times10^{-6}$ \\
 \hline
Nonlinear ($\sim \sqrt{t}$)
 & $31.1$   & $1.0\times10^2$  & $2.9\times10^{-5}$   \\
    \bottomrule
    \end{tabular}
 \end{threeparttable}
\end{table*}

\section{Summary}

As two sides of the same coin, 
turbulent reconnection and turbulent dynamo are two fundamental components of MHD turbulence. 
They result in the 
energy exchange between turbulence and magnetic fields. 
For the critical balanced MHD turbulence 
\cite{GS95}
generated in nonlinear turbulent dynamo, 
there is a balance between the dynamo growth and RD of magnetic fields at all length scales of the MHD turbulence. 
Therefore, RD of magnetic fields is indispensable for studying the nonlinear turbulent dynamo. 
It dominates over all other diffusion effects arising from plasma processes
and accounts for the inefficiency of nonlinear dynamo, as well as the large correlation length of amplified magnetic fields.

By extending the nonlinear dynamo theory to include the gravitational compression, 
we see that both turbulence and compression contribute to the growth of magnetic fields, 
and the latter becomes the dominant process with the increase of density. 
There is a weak dependence of magnetic energy on density during contraction. 
This significant deviation from the flux-freezing scaling between magnetic energy and density is attributed to the RD
in ideal MHD turbulence. 
The breakdown of the flux freezing is important for understanding the magnetic flux problem and its implications for the 
(primordial) star formation.

In a partially ionized plasma, the ionization fraction and the two-fluid (neutrals and ions)
effect should be taken into account when studying turbulent dynamo. 
When neutrals and ions are weakly coupled in a weakly ionized plasma, 
turbulent dynamo has a unique damping stage with significant ion-neutral collisional damping and a large cutoff scale of magnetic energy spectrum. 
Due to the weak neutral-ion coupling, the growing magnetic energy mainly comes from the turbulent energy carried by ions, 
which is a small fraction of the total turbulent energy contained in two fluids. 
XL16 analytical predictions on the damping stage of dynamo are numerically tested with a two-fluid dynamo simulation, 
which demonstrates the linear-in-time growth of magnetic field strength and the increasing ion-neutral collisional 
damping scale as the correlation length of magnetic fields.

Depending on the ranges of plasma parameters including magnetic Prandtl number and ionization fraction, 
the driving condition of turbulence, and the initial magnetic energy, 
turbulent dynamo can have different physical regimes and multiple evolutionary stages, 
and thus the XL16 dynamo theory can be applied to a variety of astrophysical systems. 
For instance, the nonlinear dynamo can account for the magnetic field evolution and distribution in the postshock region of a SNR, 
which explains the strong magnetic fields seen 
in shock simulations and indicated by X-ray observations. 
The damping stage of dynamo in the partially ionized preshock region serves as an alternative mechanism for magnetic field amplification 
apart from the CR-driven instabilities. 
The dynamo-amplified magnetic fields, which are also damped by ion-neutral collisional damping, 
can significantly affect the diffusion and thus the confinement of CRs at the shock. 
As the preshock dynamo depends on the upstream interstellar environment, the strength and correlation length of the amplified magnetic fields, 
and the maximum energy of CRs that can be confined by the magnetic fields can 
also have dependence on the local interstellar environment, which varies significantly in different interstellar phases.

Given the limited numerical resolution, it is very challenging to realistically simulate the astrophysical dynamo 
with an extended inertial range of turbulence and in diverse plasma conditions. 
This can result in 
discrepancies between theoretical predictions and simulations on the dynamo behavior and magnetic field structure. 
For instance, the peak scale of magnetic energy spectrum, which was believed to be close to the resistive scale based on 
earlier low-resolution dynamo simulations with insufficient turbulent inertial range 
(e.g., \cite{Mar04}),
has been shown to be significantly distant from the resistive scale by higher-resolution dynamo simulations
(e.g., \cite{Hau04}).


%
 \section*{Conflict of interest}
 The authors declare that they have no conflict of interest.

\bibliographystyle{spphys}       
\bibliography{xu}

%
%

\end{document}